\documentclass[10pt,journal,compsoc]{IEEEtran}
\ifCLASSOPTIONcompsoc
  \usepackage[nocompress]{cite}
\else
  \usepackage{cite}
\fi
\ifCLASSINFOpdf
\else
\fi
\usepackage{amsmath}
\usepackage{acronym}
\usepackage{array}
\usepackage{mdwmath}
\usepackage{mdwtab}
\usepackage{eqparbox}
\usepackage{url}
\usepackage[normalem]{ulem}

\usepackage{amsmath,amssymb,amsfonts}

\usepackage{tabularx}
\usepackage{mathtools}
\usepackage{booktabs}
\usepackage{balance}
\usepackage{xcolor}
\usepackage{graphicx}
\usepackage{textcomp}
\usepackage{subcaption}
\usepackage{wrapfig}
\usepackage{cleveref}
\usepackage{mwe}

\DeclareMathOperator*{\argmin}{arg\,min}
\usepackage[normalem]{ulem}
\begin{document}
%
% paper title
% Titles are generally capitalized except for words such as a, an, and, as,
% at, but, by, for, in, nor, of, on, or, the, to and up, which are usually
% not capitalized unless they are the first or last word of the title.
% Linebreaks \\ can be used within to get better formatting as desired.
% Do not put math or special symbols in the title.
%\title{Defensive Approximation: Securing CNNs using Approximate Computing}
\title{Defending with Errors: Approximate Computing for Robustness of Deep Neural Networks}

%
%
% author names and IEEE memberships
% note positions of commas and nonbreaking spaces ( ~ ) LaTeX will not break
% a structure at a ~ so this keeps an author's name from being broken across
% two lines.
% use \thanks{} to gain access to the first footnote area
% a separate \thanks must be used for each paragraph as LaTeX2e's \thanks
% was not built to handle multiple paragraphs
%
%
%\IEEEcompsocitemizethanks is a special \thanks that produces the bulleted
% lists the Computer Society journals use for "first footnote" author
% affiliations. Use \IEEEcompsocthanksitem which works much like \item
% for each affiliation group. When not in compsoc mode,
% \IEEEcompsocitemizethanks becomes like \thanks and
% \IEEEcompsocthanksitem becomes a line break with idention. This
% facilitates dual compilation, although admittedly the differences in the
% desired content of \author between the different types of papers makes a
% one-size-fits-all approach a daunting prospect. For instance, compsoc 
% journal papers have the author affiliations above the "Manuscript
% received ..."  text while in non-compsoc journals this is reversed. Sigh.

\author{Amira~Guesmi,~%\IEEEmembership{Member,~IEEE,}
        Ihsen~Alouani,~%\IEEEmembership{Fellow,~OSA,}
        Khaled~N.~Khasawneh,~%\IEEEmembership{Fellow,~OSA,}
        Mouna~Baklouti,~%\IEEEmembership{Fellow,~OSA,}
        Tarek~Frikha,~\\
        Mohamed~Abid,~
        and~Nael~Abu-Ghazaleh~%\IEEEmembership{Life~Fellow,~IEEE}% <-this % stops a space

\IEEEcompsocitemizethanks{\IEEEcompsocthanksitem A. Guesmi, M. Baklouti, T. Frikha and M. Abid are with ENIS, University of Sfax, Tunisia.\protect\\
%GA, 30332.\protect\\
% note need leading \protect in front of \\ to get a newline within \thanks as
% \\ is fragile and will error, could use \hfil\break instead.
\IEEEcompsocthanksitem I. Alouani is with IEMN, Polytechnic University Hauts-De-France, France.\protect\\E-mail: ihsen.alouani@uphf.fr %see http://www.michaelshell.org/contact.html
\IEEEcompsocthanksitem K. N. Khasawneh is with George Mason University, USA.\protect\\\IEEEcompsocthanksitem N. Abu-Ghazaleh is with University of California Riverside, USA.}% <-this % stops a space
%\thanks{Manuscript received April 19, 2005; revised August 26, 2015.}
}

% note the % following the last \IEEEmembership and also \thanks - 
% these prevent an unwanted space from occurring between the last author name
% and the end of the author line. i.e., if you had this:
% 
% \author{....lastname \thanks{...} \thanks{...} }
%                     ^------------^------------^----Do not want these spaces!
%
% a space would be appended to the last name and could cause every name on that
% line to be shifted left slightly. This is one of those "LaTeX things". For
% instance, "\textbf{A} \textbf{B}" will typeset as "A B" not "AB". To get
% "AB" then you have to do: "\textbf{A}\textbf{B}"
% \thanks is no different in this regard, so shield the last } of each \thanks
% that ends a line with a % and do not let a space in before the next \thanks.
% Spaces after \IEEEmembership other than the last one are OK (and needed) as
% you are supposed to have spaces between the names. For what it is worth,
% this is a minor point as most people would not even notice if the said evil
% space somehow managed to creep in.

% The paper headers
\markboth{Journal of \LaTeX\ Class Files,~Vol.~14, No.~8, August~2015}%
{Shell \MakeLowercase{\textit{et al.}}: Bare Advanced Demo of IEEEtran.cls for IEEE Computer Society Journals}
% The only time the second header will appear is for the odd numbered pages
% after the title page when using the twoside option.
% 
% *** Note that you probably will NOT want to include the author's ***
% *** name in the headers of peer review papers.                   ***
% You can use \ifCLASSOPTIONpeerreview for conditional compilation here if
% you desire.

% The publisher's ID mark at the bottom of the page is less important with
% Computer Society journal papers as those publications place the marks
% outside of the main text columns and, therefore, unlike regular IEEE
% journals, the available text space is not reduced by their presence.
% If you want to put a publisher's ID mark on the page you can do it like
% this:
%\IEEEpubid{0000--0000/00\$00.00~\copyright~2015 IEEE}
% or like this to get the Computer Society new two part style.
%\IEEEpubid{\makebox[\columnwidth]{\hfill 0000--0000/00/\$00.00~\copyright~2015 IEEE}%
%\hspace{\columnsep}\makebox[\columnwidth]{Published by the IEEE Computer Society\hfill}}
% Remember, if you use this you must call \IEEEpubidadjcol in the second
% column for its text to clear the IEEEpubid mark (Computer Society journal
% papers don't need this extra clearance.)

% use for special paper notices
%\IEEEspecialpapernotice{(Invited Paper)}

% for Computer Society papers, we must declare the abstract and index terms
% PRIOR to the title within the \IEEEtitleabstractindextext IEEEtran
% command as these need to go into the title area created by \maketitle.
% As a general rule, do not put math, special symbols or citations
% in the abstract or keywords.
\IEEEtitleabstractindextext{%
\begin{abstract}
%In the past few years, an increasing number of machine-learning architectures, such as Convolutional Neural Networks (CNNs), have been applied to solving a wide range of real-life problems.  
 Machine-learning architectures, such as Convolutional Neural Networks (CNNs) are vulnerable to adversarial attacks: inputs crafted carefully to force the system output to a wrong label. Since machine-learning is being deployed in safety-critical and security-sensitive domains, such attacks may have catastrophic security and safety consequences.  
In this paper, we propose for the first time to use hardware-supported approximate computing to improve the robustness of machine-learning classifiers. 
%We show that our approximate computing implementation achieves robustness across a wide range of attack scenarios. Specifically, we show that successful adversarial attacks against the exact classifier have poor transferability to the approximate implementation. The transferability is even poorer for the black-box attack scenarios, where adversarial attacks are generated using a proxy model. 
we show that successful adversarial attacks against the exact classifier have poor transferability to the approximate implementation.
Surprisingly, the robustness advantages also apply to white-box attacks where the attacker has unrestricted access to the approximate classifier implementation: in this case, we show that substantially higher levels of adversarial noise are needed to produce adversarial examples. 
Furthermore, our approximate computing model maintains the same level in terms of classification accuracy, does not require retraining, and reduces resource utilization and energy consumption of the CNN.
We conducted extensive experiments on a set of strong adversarial attacks; We empirically show that the proposed implementation increases the robustness of a LeNet-5, Alexnet and VGG-11 CNNs considerably with up to $50\%$ by-product saving in energy consumption due to the simpler nature of the approximate logic. %up to $99\%$ and $87\%$, respectively for strong transferability-based attacks along % %We also show that a white-box attack requires a remarkably higher noise budget to fool the approximate classifier, causing an average of $4~dB$ degradation of the PSNR of the input image relative to the images that succeed in fooling the exact classifier.
%We also show that for a white-box setting, DA increases model resilience to reach 90\% for images from CIFAR-10 and ImageNet under the PGD attack for high noise budget while maintaining a minor impact on model performance on clean images.

\end{abstract}

% Note that keywords are not normally used for peerreview papers.
}

% make the title area
\maketitle

% To allow for easy dual compilation without having to reenter the
% abstract/keywords data, the \IEEEtitleabstractindextext text will
% not be used in maketitle, but will appear (i.e., to be "transported")
% here as \IEEEdisplaynontitleabstractindextext when compsoc mode
% is not selected <OR> if conference mode is selected - because compsoc
% conference papers position the abstract like regular (non-compsoc)
% papers do!
\IEEEdisplaynontitleabstractindextext
% \IEEEdisplaynontitleabstractindextext has no effect when using
% compsoc under a non-conference mode.

% For peer review papers, you can put extra information on the cover
% page as needed:
% \ifCLASSOPTIONpeerreview
% \begin{center} \bfseries EDICS Category: 3-BBND \end{center}
% \fi
%
% For peerreview papers, this IEEEtran command inserts a page break and
% creates the second title. It will be ignored for other modes.
\IEEEpeerreviewmaketitle

\ifCLASSOPTIONcompsoc
\IEEEraisesectionheading{\section{Introduction}\label{sec:introduction}}
\else
\section{Introduction}
\label{sec:introduction}
\fi
\IEEEPARstart{C}{onvolutional} neural networks (CNNs) and other deep learning structures  provide state-of-the-art performance in many application domains, such as computer vision %\cite{simonyan2014deep,redmon2016yolo9000}
, natural language processing %\cite{deng2018deep}
, robotics %\cite{pierson2017deep}
, autonomous driving %\cite{al2017deep} 
, and healthcare %\cite{miotto2018deep}
. With the rapid progress in CNN's development and deployment, they are increasing concerns about their vulnerability to adversarial attacks: maliciously designed imperceptible perturbations injected within the data that cause CNNs to misclassify.   Adversarial attacks have been demonstrated in real-world scenarios \cite {phys%,phy9,phy10,neuroattack
}, making this vulnerability a serious threat to safety-critical and other applications that rely on CNNs. 

Several software-based defenses have been proposed against Adversarial Machine Learning (AML) attacks~\cite{distillation_SP, na2017cascade}, but more advanced attack strategies~\cite{C&W, pgd} also continue to evolve that demonstrate vulnerability of some of these defenses. Moreover, many of the proposed defenses introduce substantial overheads to either the training or the inference operation of CNNs~\cite{ na2017cascade}.  These overheads increase the computational requirements of these systems, which is already a significant challenge driving substantial research into algorithmic and hardware techniques to improve CNN performance and energy-efficiency~\cite{reviewDL}.   
Thus, finding new approaches to harden systems against AML without heavy overheads in both design-time and run-time is an area of substantial need.

In this paper, we propose a new hardware based approach to improve the robustness of machine learning (ML) classifiers.  Specifically, we show that Approximate Computing (AC),  designed to improve the performance and power consumption of algorithms and processing elements,  can substantially improve CNN robustness to AML.  
Our technique, which we call \emph{defensive approximation} (DA), substantially enhances the robustness of CNNs to adversarial attacks. %Note that, throughout the paper, we refer to classifiers that use DA as approximate classifiers.  
We show that for a variety of attack scenarios, and utilizing a range of algorithms for generating adversarial attacks, DA provides substantial robustness even under the assumptions of a powerful attacker with full access to the internal classifier structure.   Importantly, DA does not require retraining or fine-tuning, allowing pre-trained models to benefit from its robustness and performance advantages by simply replacing the exact multiplier implementations with approximate ones.   The approximate classifier achieves similar accuracy to the exact classifier for Lenet-5 and Alexnet. 

DA also benefits from the conventional advantages of AC, resulting in a less complex design that is both faster and more energy efficient.   Other defenses such as Defensive Quantization (DQ)~\cite{DQ} also provide energy efficiency benefits in addition to robustness.  However, we show that because it is input-dependent DA achieves twice higher robustness to attacks than DQ.

We carry out several experiments to better understand the robustness advantages of DA.  We show that the unpredictable and input-dependent variations introduced by AC improve the CNN resilience to adversarial perturbations. Experimental results show that DA has a confidence enhancement impact on non-adversarial examples; we believe that this is due to our AC multiplier which adds input dependent approximation with generally higher magnitude at large multiplication values. In fact, the AC-induced noise in the convolution layer is shown to be higher in absolute value when the input matrix is highly correlated to the convolution filter, and by consequence highlights further the features. This observation at the feature map propagates through the model and results in enhanced classification confidence, i.e., the difference between the $1^{st}$ class and the "runner-up". Intuitively and as shown by prior work~\cite{smooth}, enhancing the confidence furthers the classifier's robustness.
At the same time, we observe negligible accuracy loss compared to a conventional CNN implementation on non-adversarial inputs while providing considerable power savings.

In summary, the contributions of the paper are:
\begin{itemize}
    \item We build an aggressively approximate floating point multiplier that injects data-dependent noise within the convolution calculation. Subsequently, we used this approximate multiplier to implement an approximate CNN hardware accelerator (Section \ref{sec:AppxCNN}).  
    \item To the best of our knowledge, we are the first to leverage AC to enhance CNN robustness to adversarial attacks without the need for re-training, fine-tuning, nor input pre-processing. We investigate the capacity of AC to help defending against adversarial attacks in Section \ref{sec:secu}.
    \item We empirically show that the proposed approximate implementation reduces the success rate of adversarial attacks under grey-box setting by an average of 87\% and 71.5\% in Lenet-5 and Alexnet CNNs respectively. 
    For a white-box setting, DA is found to help increase model resilience to reach 90\% for images from CIFAR-10 and ImageNet under the PGD attack for a noise budget of $0.06$.  
    \item The proposed technique has minor impact on model performance on clean images.
    %\ihsen{Could you please update the results with the new benchmarks}
    %\item We evaluate the approximate classifiers against powerful attackers with white-box access.  We observe that attackers require substantially higher adversarial perturbations to fool the approximate classifier.
    \item We propose an approximate BFloat16 multiplier that stills perform well in spite of the aggressive approximation. %\ihsen{some contribution about approximate Cov or/and FC }
    \item We study the impact of introducing approximation to convolution and fully-connected (FC) layers on model accuracy.
    \item We build an approximate BFloat16 multiplier which resulted in further gain in energy by up to 81\% and 88\% saving in delay compared to a conventional FP multiplier. %\ihsen{compared to ?}.
    \item \textbf{Open source:} for reproducible research, we release the complete source code of our method. \footnote{https://github.com/AG-X09/Defensive-Approximation}

\end{itemize}

\noindent 
We believe that DA is highly practical; it can be deployed without retraining or fine-tuning, achieving comparable classification performance to exact classifiers.  In addition to security advantages, DA {\em improves} performance by reducing latency and energy making it an attractive choice even in Edge device settings. % (Appendix \ref{sec:perf_impl})

An earlier version of this paper appeared in the International Conference on Architectural Support for Programming Languages and Operating Systems (ASPLOS) \cite{DA}.\footnote{The description of the the contributions of the journal extension vs. the conference article is included in the cover letter.}  %We will remove it from the paper should it be accepted for publication.

\section{Background}
This section first presents an overview of adversarial attacks followed by introducing AC.  

\subsection{Adversarial Attacks}

Deep learning techniques gained popularity in recent years and are now deployed even in safety-critical tasks, such as recognizing road signs for autonomous vehicles \cite{signs}. Despite their effectiveness and popularity, CNN-powered applications are facing a critical challenge – adversarial attacks. Many studies \cite{vulnerable, fgsm} have shown that CNNs are vulnerable to carefully crafted inputs designed to fool them, very small imperceptible perturbations added to the data can completely change the output of the model. In computer vision domain, these adversarial examples are intentionally generated images that look almost exactly the same as the original images, but can mislead the classifier to provide wrong prediction outputs. Other work \cite{noneed} claimed that adversarial examples are not a practical threat to ML in real-life scenarios. However, physical adversarial attacks have recently been shown to be effective against CNN based applications in real-world \cite{phys}.

\noindent 
{\bf Minimizing Injected Noise:} Its essential for the adversary to minimize the added noise to avoid detection.  
%, an adversary, using information learned about the structure of the classifier, tries to craft perturbations added to the input to cause incorrect classification.   In these types of attacks, the adversary desires to minimize this perturbation noise to avoid detection.  
For illustration purposes, consider a CNN used for image classification.  More formally, given an original input image $x$ and a target classification model $ f() ~s.t. ~ f(x) = l $, the problem of generating an adversarial example $x^*$ can be formulated as a constrained optimization \cite{pbform}:

\begin{equation}
\label{eq:adv}
     \begin{array}{rlclcl}
        x^* = \displaystyle \argmin_{x^*}  \mathcal{D}(x,x^*),  
         s.t.  ~ f(x^*) = l^*,  ~ l \neq l^*
\end{array}
\end{equation}

Where $\mathcal{D}$ is the distance metric used to quantify the similarity between two images and the goal of the optimization is to minimize this added noise, typically to avoid detection of the adversarial perturbations. $l$ and $l^*$ are the two labels of $x$ and $x^*$, respectively:  $x^*$ is considered as an adversarial example if and only if the label of the two images are different ($ f(x) \neq  f(x^*) $) and the added noise is bounded ($\mathcal{D}(x,x^*) < \epsilon $ where $\epsilon \geqslant 0 $).

%\noindent
%{\bf Distance Metrics:}
%The adversarial examples and the added perturbations should be visually imperceptible by humans. Since it is hard to model humans' perception, researchers proposed three metrics to approximate humans' perception of visual difference, namely $L_0$, $L_2$, and $L_ \infty$ \cite{C&W}. These metrics are special cases of the $L_p$ norm: 
%\begin{equation}
%    \left\|x\right\|_p = \left( \sum^{n}_{i = 1} \left |x_i \right | ^{p} \right)^{\frac{1}{p}}
%\end{equation}
%These three metrics focus on different aspects of visual significance. $L_0$ counts the number of pixels with different values at corresponding positions in the two images. $L_2$ measures the Euclidean distance between the two images $x$ and $x^*$. $L_ \infty$ measures the maximum difference for all pixels at corresponding positions in the two images.

%The consequences of an adversarial attack can be dramatic.  For example, misclassification of a stop sign as a yield sign or a speed limit sign could lead to material and human damages. Another possible situation is when using CNNs in financial transactions and automatic bank check processing -- using handwritten character recognition algorithms to read digits from bank cheques or using neural networks for amount and signature recognition \cite{cheque}. An attacker could easily fool the model to predict wrong bank account numbers or amount of money or even fake a signature. A dangerous situation, especially with such large sums of money at stake.

\subsection{Approximate Computing}

The speed of new generations of computing systems, from embedded and mobile devices to servers and computing data centers, has been drastically climbing in the past decades. This development was made possible by the advances in integrated circuits (ICs) design and driven by the increasingly high demand for performance in the majority of modern applications. However, this development is physically reaching the end of Moore’s law, since TSMC and Samsung are releasing 5 $nm$ technology \cite{tsmc_moore}. On the other hand, a wide range of modern applications is inherently fault-tolerant and may not require the highest accuracy. This observation has motivated the development of approximate computing (AC), a computing paradigm that trades power consumption with accuracy. The idea is to implement inexact/AC elements that consume less energy, as far as the overall application tolerates the imprecision level in computation.
This paradigm has been shown promising for inherently fault-tolerant applications such as deep/ML, big data analytics, and signal processing. Several AC techniques have been proposed in the literature and can be classified into three main categories based on the computing stack layer they target: software, architecture, and circuit level~\cite{Survey}. 

In this paper, we consider AC for a \emph{totally new objective}; enhancing CNNs robustness to adversarial attacks, without losing the initial advantages of AC.

\section{Threat Model}
We assume an attacker attempting to conduct adversarial attacks to fool a classifier in a variety of attack scenarios.%, which we overview in this section.  

\subsection{Adversary Knowledge (Attacks Scenarios)}

In this work, we consider three attack scenarios:

\noindent
    \textbf{\textit{Transferability Attack.}} We assume the adversary is aware of the exact classifier internal model; its architecture and parameters. The adversary uses the exact classifier to create adversarial examples. Thus, we explore whether these examples transfer effectively to the approximate classifier (DA classifier).
    
  \noindent  
    \textbf{\textit{Black-box Attack.}} We assume the attacker has access only to the input/output of the victim classifier (which is our approximate classifier) and has no information about its internal architecture. The adversary first uses the results of querying the victim to reverse engineer the classifier and create a substitute %\khaled{proxy or substitute?} 
    CNN model. With the substitute model, the attacker can attempt to generate different adversarial examples to attack the victim classifier.

\noindent
    \textbf{\textit{White-box Attack.}} We assume a powerful attacker who has full knowledge of the victim classifier's architecture and parameters (including the fact that it uses AC).  The attacker uses this knowledge to create adversarial examples.  % and the defense strategy used which means that the adversary is capable of generating adversarial examples targeting the victim classifier.  

\subsection{Adversarial Example Generation}

 We consider several adversarial attack generation algorithms for our attack scenarios, including some of the most recent and potent evasion attacks.  Generally, in each algorithm, the attacker tries to evade the system by adjusting malicious samples during the inference phase, assuming no influence over the training data. However, as different defenses have started to be deployed that specifically target individual adversarial attack generation strategies, new algorithms have started to be deployed that bypass these defenses. For example, methods such as defensive distillation \cite{distillation_SP} were introduced and demonstrate robustness against the FGSM attack %\khaled{what is this attack?}
 \cite{fgsm}.  However, the new $C\&W$ attack was able to bypass these defenses \cite{C&W}.  
Thus, demonstrating robustness against a range of these attacks provides confidence that a defense is effective in general, against all known attack strategies, rather than against a specific strategy.

These attacks can be divided into three categories: Gradient-based attacks relying on detailed model information, including the gradient of the loss w.r.t. the input. Score-based attacks rely on the predicted scores, such as class probabilities or \textit{logits} of the model. On a conceptual level, these attacks use the predictions to estimate the gradient numerically.  Finally, decision-based attacks rely only on the final output of the model and minimizing the adversarial examples' norm.  %The attacks are summarized in Table \ref{Attack_methods}. 

% \begin{table}[!htp]
% \small
% \centering
%   \caption{Summary of the used attack methods. Notice that the strength estimation is based on~\cite{strength}.} 
%   \label{Attack_methods}
%   \begin{tabular}{ccccc}
%     \toprule
%     \textbf{Method} & \textbf{Category}  & \textbf{Norm} & \textbf{Learning} & \textbf{Strength} \\
%     \midrule
%     FGSM  & gradient-based   &  $L_\infty$  & One shot  & ***\\
%     PGD   & gradient-based   &  $L_\infty$  & Iterative & ****\\
%     JSMA  & gradient-based   &  $L_0$       & Iterative & ***\\
%     C\&W & gradient-based   &  $L_2$       & Iterative & *****\\
%     DF  & gradient-based   &  $L_2$       & Iterative & ****\\
%     LSA  & Score-based      &  $L_2$       & Iterative & ***\\
%     BA & Decision-based   &  $L_2$       & Iterative & ***\\
%     HSJ   & Decision-based   &  $L_2$       & Iterative & *****\\
% %    Universal perturbations & & Untargeted & Universal & $L_2$ & Iterative & *****\\
    
%   \bottomrule
% \end{tabular}
% \end{table}

\label{sec:threat}

\section{Defensive Approximation: Implementing Approximate CNNs}
%\subsection{Approximate CNN}

We propose to leverage approximate computing to improve the robustness of ML classifiers, such as CNNs, against adversarial attacks. We call this general approach \textbf{Defensive Approximation} (DA). The closest approach to DA is the perturbation-based defense~\cite{snP2019_certif,smooth} that either adds noise or otherwise filter the input data to try to interfere with any adversarial modifications to the input of a classifier. However, our approach advances the state-of-the-art by injecting perturbations throughout the classifier and directly by the approximate hardware, thereby enhancing both robustness and power efficiency. 

Technically, the approximate design process is driven by two main considerations:
\begin{enumerate} %[label=(\alph*)]
    \item Injecting significant noise that can influence the CNN behavior, and allows by-product power gains. %This led us to choose an aggressively approximate multiplier to be deployed.
    \item Keeping the cumulative noise magnitude under control to avoid impacting the baseline accuracy of the CNN. %For this reason, we purpose to change the mantissa multiplier only (fraction), while keeping the exponent part unchanged (exact).
\end{enumerate}

For consideration (b), we purpose to an implementation that replaces the conventional mantissa multiplier in floating point multipliers, with a simpler approximate design. This choice is backed by the study in \cite{ D&T}, which show that errors in the exponent part might have a drastic impact on CNNs accuracy. 
Consideration (a) means that the approximate design needs to be aggressive to induce significant noise. Therefore, as detailed in the next subsection, we chose the corner case, i.e., the most aggressive approximate design \cite{AMA5, Heap}, and used it to replace the mantissa computation logic in a basic floating point multiplier. %Moreover, for the sake of comparison, we also proceed to a design space exploration and compare our design to a less aggressive approximate design \cite{Heap}.} 

In this section, we present our approximate multiplier design and analyze its properties.
%\nael{I think this opens a door for the reviewers to question our time and power analysis.   Are you evaluating the mantissa manipulation and the rounding logic in the power and latency analysis?} \ihsen{The power and delay analysis is for the Mantissa Multiplier, and I think I can have numbers for the whole FP multiplier.}

\subsection{Approximate Floating Point Multiplier}
\label{trend}
%\textcolor{blue}{Khaled: we do not discuss HEAP anywhere in this subsection}
ML structures, such as CNNs, often rely on computationally expensive operations, e.g., convolutions that are composed of multiplications and additions. Floating-point multiplications consume most of the processing energy in both inference and training of CNNs~\cite{Heap}. 
Although approximate computation can be introduced in different ways (with likely different robustness benefits), DA leverages a new approximate 32-bit floating-point multiplier, which we call \textit{approximate floating-point multiplier} (Ax-FPM).
The IEEE 754-2008 compliant floating-point format binary numbers are composed of three parts: a sign, an exponent, and a mantissa (also called fraction)~\cite{FP}. %\nael{note missing citation} 
The sign is the most significant bit, indicating whether the number is positive or negative. In a single-precision format, the following $8$ bits represent the exponent of the binary number ranging from $-126$ to $127$. The remaining $23$ bits represent the fractional part (mantissa). 
For most of the floating number range, the normalized format is:
\begin{equation}\label{eqn:ieee754}
val = ( -1 )^{\textit{sign}} \times 2^{exp-bias} \times ( 1.fraction ) 
\end{equation}

A floating-point multiplier (FPM) consists mainly of three units: mantissa multiplier, exponent adder, and a rounding unit.  The mantissa multiplication consumes $81\%$ of the overall power of the multiplier \cite{Heap}.

Ax-FPM is designed based on a mantissa multiplication unit that is constructed using approximate full adders (FA). The FAs are aggressively approximated to inject computational noise within the circuit. We describe Ax-FPM by first presenting the approximate FA design, and then the approximate mantissa multiplier used to build the Ax-FPM.

\begin{figure}[!htp]
\centering
\includegraphics[width=\columnwidth]{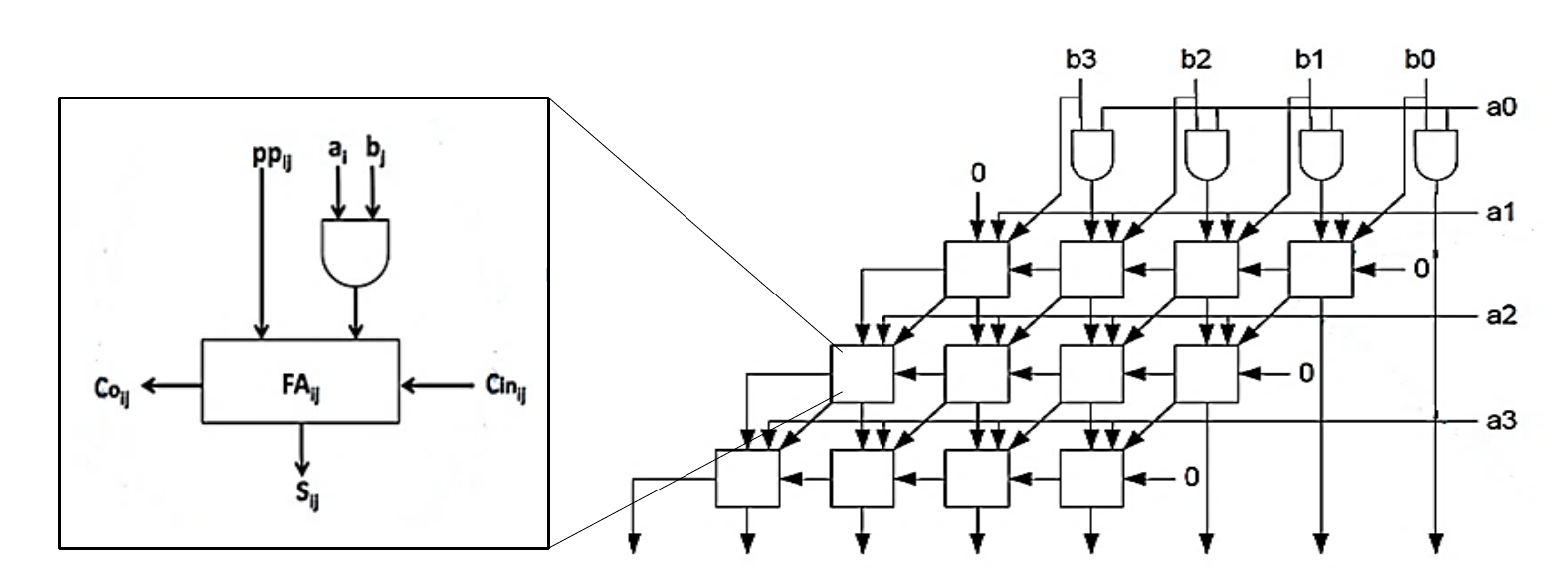}
\caption{An illustration of a $4 \times 4$ array multiplier architecture.}
\label{MA}
\end{figure}

\begin{figure}[!htp]
\centering
\includegraphics[width=\columnwidth]{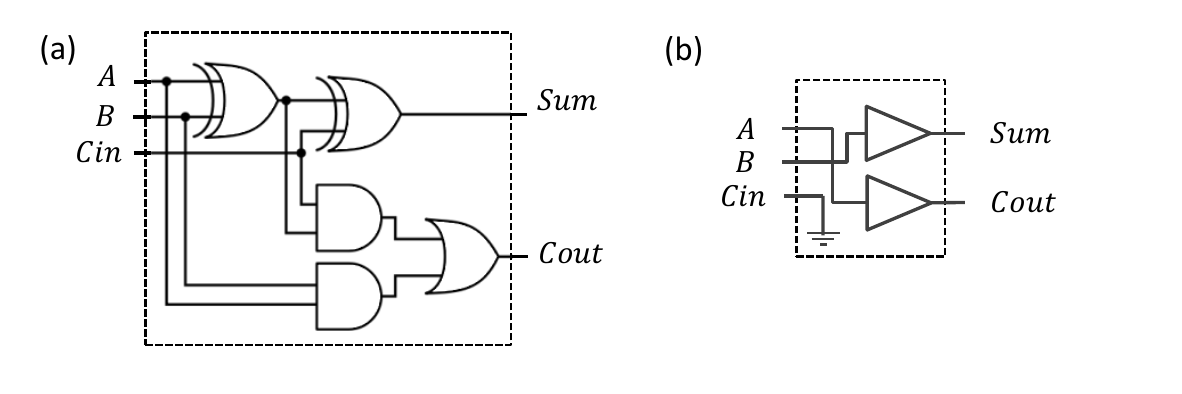}
\caption{Logic diagram of (a) exact Full Adder, (b) AMA5 (used in this work).}
\label{FA}
\end{figure}

To build a power-efficient and a higher performance FPM, we propose to replace the mantissa multiplier by an approximate mantissa multiplier; an array multiplier constructed using approximate FAs. We selected an array multiplier implementation because it is considered one of the most power-efficient among conventional multiplier architectures~\cite{array}.  In the array architecture, multiplication is implemented through the addition of partial products generated by multiplying the multiplicand with each bit of multiplier using AND gates, as shown in Figure \ref{FA}.  %In the case of a 32-bit floating-point multiplier, the mantissa multiplication is a $24\times24$-bit multiplication. %The exact full adder was replaced by an approximate mirror adder (AMA5). 

Specifically, we build an array multiplier based on an approximate mirror adder (AMA5)~\cite{AMA5} in place of exact FAs. The approximation of a conventional FA is performed by removing some internal circuitry, thereby resulting in power and resource reduction at the cost of introducing errors.  Consider a FA with (A,B, $C_{in}$) as inputs and ($Sum$, $C_{out}$) as outputs ($C$ here refers to carry). For any input combination, the logical \textit{approximate} expressions for $Sum$ and $C_{out}$ are: $Sum = B$ and $C_{out} = A$. 
The AMA5 design is constituted by only two buffers (see Figure \ref{FA}), leading to the latency and energy savings relative to the exact design, but more importantly, introduce errors into the computation.  It is worth noting that these errors are data dependent, appearing for specific combinations of the inputs, and ignoring the carry in value, making the injected noise difficult to predict or model.  

\begin{figure}[!htp]
\centering
\includegraphics[width=0.8\columnwidth]{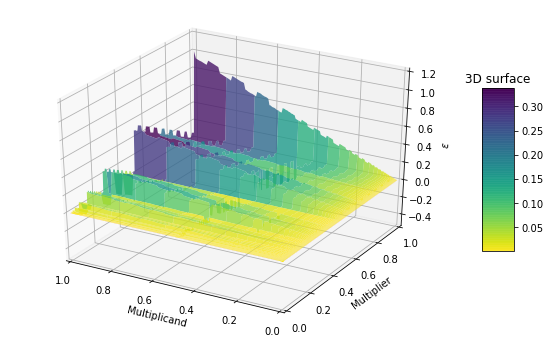}
\caption{Noise introduced by the approximate multiplier while the operands in $[0,1]$.
}
\label{appx_mult}
\end{figure}

When trying to evaluate the proposed Ax-FPM, we were interested in studying its behavior under small input numbers ranging between $-1$ and $+1$ since most of the internal operations within CNNs are in this range.
We measure the introduced error as the difference of the output of the approximate multiplier and the exact multiplier. The results are shown in Figure \ref{appx_mult} using 100 million randomly generated multiplications across the input range from $-1$ to $1$. 

Three trends can be observed that will be used later to help understanding the impact of the approximation using Ax-FPM on CNN security: \textbf{(i)} The first is the data-dependent discontinuity of the approximation-induced errors, \textbf{(ii)}  We noticed that in $96\%$ of the cases, the approximate multiplication results in higher absolute values than the exact multiplication: For positive products, the approximate result is higher than the exact result, and for negative product results the approximate result is lower than the exact result, and \textbf{(iii)} In general, we notice that the larger the multiplied numbers, the larger the error magnitude added to the approximate result. As shown later, these observations will help understand the mechanism that we think is behind robustness enhancement.

\begin{figure}[!htp]
\centering
\includegraphics[width=3.5in]{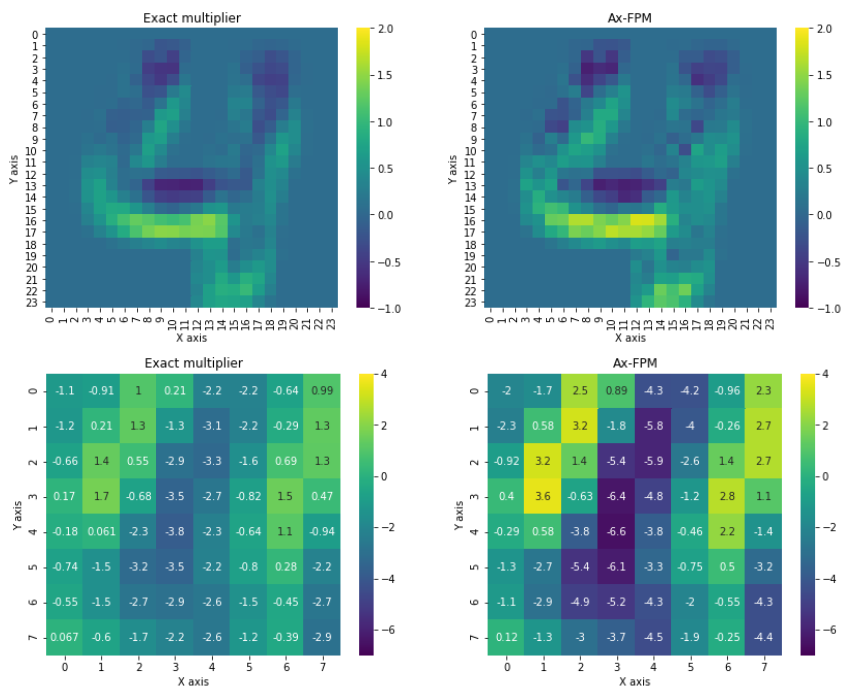}
\caption{Convolution outputs.}
\label{fig_sim}
\end{figure}

%\begin{figure}[!htp]
%\centering
%\includegraphics[width=0.9\columnwidth,height=4cm]{figures/Convolution.pptx.pdf}
%\caption{Convolution result of the filter and each input image using exact and %approximate convolution.}
%\label{appx_conv}
%\end{figure}

\subsection{Approximate BFloat16 multiplier}
\label{appx_bfloat}
% \noindent\colorbox{Yellow}{%
%     \parbox{\dimexpr\linewidth-2\fboxsep}
% {
BFloat16 (Brain Floating Point) \cite{ kalamkar2019study} is a data representation used in machine learning applications and intended to reduce the storage requirements and increase computing speed without losing precision. In this section, we use the same previously described approach to build an approximate BFloat16 multiplier.

\begin{figure}[!htp]
\centering
\includegraphics[width=1\columnwidth]{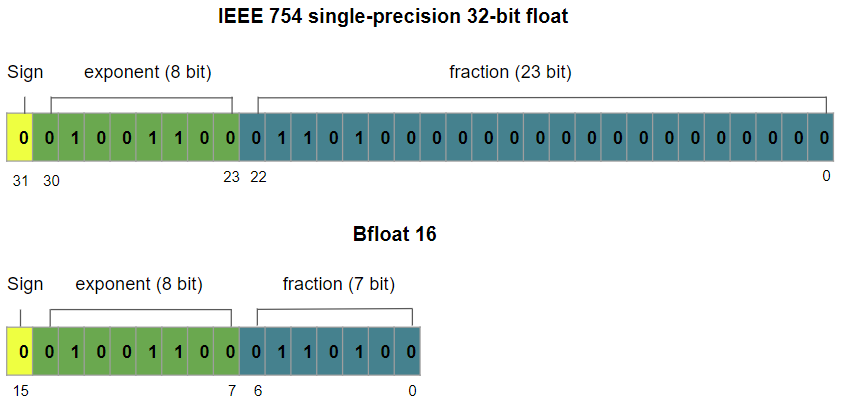}
\caption{Contrast between IEEE 754 Single-precision 32-bit floating-point format and Bfloat16 format.
}
\label{fig:bfloat}
\end{figure}

The BFloat16 is a truncated version of the 32-bit IEEE 754 single-precision floating-point format (float32). The BFloat16 format is shown in Figure \ref{fig:bfloat}; it consists of 1 sign bit, an 8-bit exponent, and a 7-bit mantissa giving the range of a full 32-bit number but in the data size of a 16-bit number. 

\begin{figure}[!htp]
\centering
\includegraphics[width=3in]{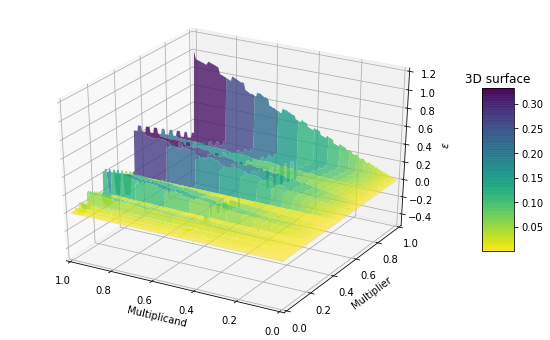}
\caption{Noise introduced by the approximate BFloat16 multiplier.}
\label{fig_bfloat}
\end{figure}

The proposed approximate BFloat16 multiplier architecture is uses an $8 \times 8$ approximate array mantissa multiplier based on AMA5 full adders. 
%We use the same approach as in the Ax-FPM and we build an approximate $8 \times 8$ mantissa multiplier. %instead of $24 \times 24$ mantissa multiplier. 
The error distribution induced by the proposed approximate BFloat16 multiplier is shown in Figure \ref{fig_bfloat}. Again, we can notice the same trends previously introduced in Section \ref{trend} for the Ax-FPM.
%}}

\subsection{Extending approximation to fully connected layers}
\label{appx_fc}

% \noindent\colorbox{Yellow}{%
%     \parbox{\dimexpr\linewidth-2\fboxsep}
% {
In order to investigate the impact of AC at larger scales than the individual multiplication, we track the impact on convolution operations and fully connected layers. The approximate CNN is built using the approximate convolution operations and approximate fully connected layers as building blocks. The activation functions and the pooling layers which do not use multiplication are similar to the conventional CNN.   
%\ihsen{$\rightarrow$ This following paragraph seems to me prematurely dropped here-- we actually do not have this intuition before hand; I suggest we move it to the section where we try to analyse and understand what happened. } \amira{not sure I got your point}
 %As we can see, the approximate model resulted in higher scores 
%}}

\begin{figure}[!htp]
\centering
\includegraphics[width=0.9\columnwidth,height=5cm]{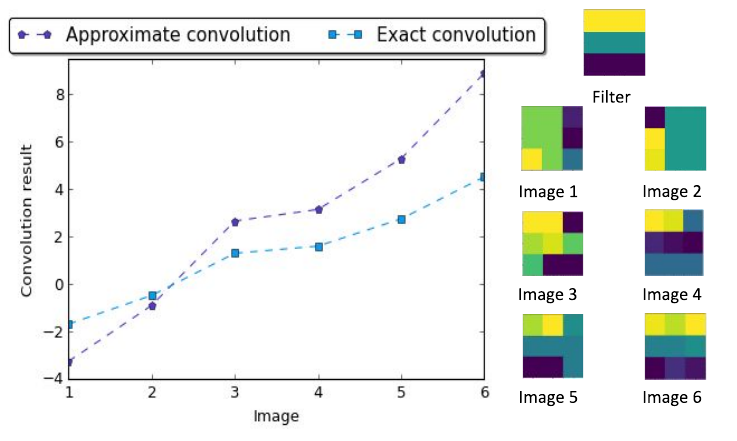}
\caption{Convolution result of the filter and each input image using exact and approximate convolution.}
\label{appx_conv}
\end{figure}

In Figure \ref{appx_conv}, we run an experiment where we choose a filter and six different images with different degrees of similarity to the chosen filter (1 to 6 from the least to the most similar), and we perform the convolution operation. We notice that the approximate convolution using Ax-FPM delivers higher results for similar inputs and lower results for dissimilar inputs. We can also notice that the higher the similarity, the higher the gap between the exact and Ax-FPM approximation results.  

Convolution layers enable CNNs to extract data-driven features rather than relying on manual feature-extraction in classical ML systems. The convolution operation is performed on the layer's input data using a kernel matrix that is convoluted (piece-wise multiplied) against the input data to produce a feature map. As we slide the filter over the input from left to right and top to bottom, whenever the filter coincides with a \textbf{similar} portion of the input, the convolution result is high, and the neuron will fire. 
The weight matrix filters out portions of the input image that does not align with the filter, and the approximate multiplier helps improve this process by further increasing the output when a feature is detected.

Therefore, by using the approximate convolution, the main features of the image that are important in the image recognition are retained and further highlighted with higher scores as illustrated in Figure \ref{fig_sim}. %that will later help increase the confidence of the classification result, as explained in Section \ref{sec:how}. 

\begin{figure}[!htp]
\centering
\includegraphics[width=3.5in]{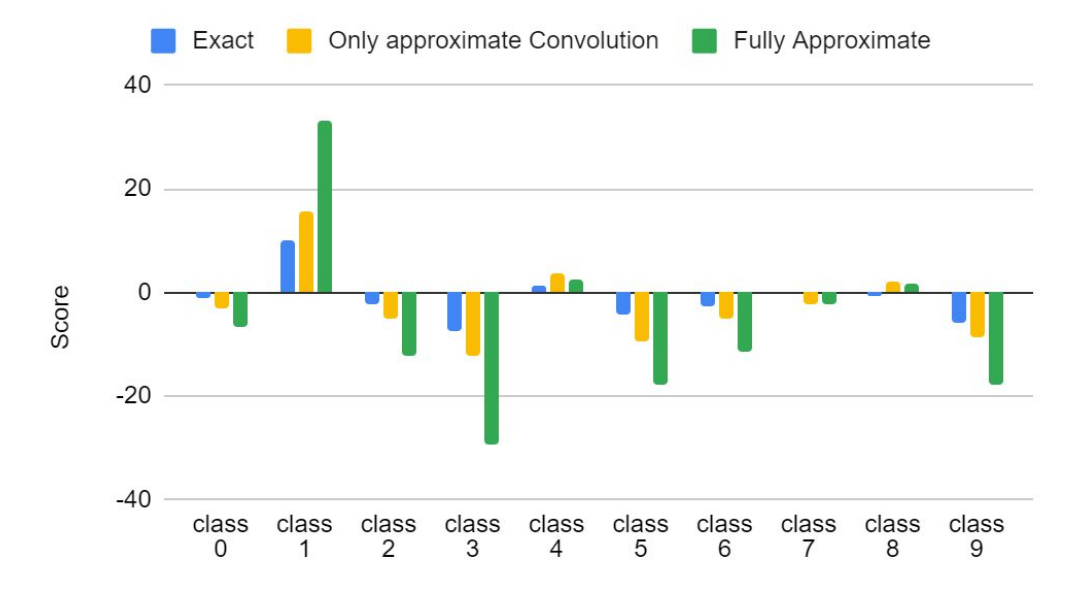}
\caption{The final raw average scores of multiple samples inference from exact, only approximate convolution bloc and fully approximate LeNet-5.}%\nael{I found this figure distracting here...}
\label{fig_out}
\end{figure}

Same with approximate fully connected (FC) layers, when using our approximate multiplier, we further increase the score of most relevant information and lower that of the less relevant ones. %Thus, final score w.r.t. to the true class as shown in Figure \ref{fig_out}.
Thus, further increasing the confidence (the output of the pre-softmax layer) of the classification result as shown in Figure \ref{fig_out},
where we select 5 random samples $\{S1, ..., S5\}$ belonging to the same class and perform inferences using the fully approximate, only approximate convolution bloc, and the conventional models. The score of each class is defined as the average of the 5 samples corresponding score.

Therefore, by using the approximate convolution, the main features of the image that are important in the image recognition are retained and further highlighted with higher scores that will later help increase the confidence of the classification result, as explained in Section \ref{sec:how}.
%\subsection{AC Design Space Exploration}
%For the sake of comparison, we proceed to a design space exploration considering the accuracy of the multipliers and resource utilization as optimization objectives. The comparison of our design with the optimal approximate design given by the exploration and referred to as HEAP led to a clearly dominant choice in favor of our design given the following perspectives:

%\textbf{(i) Resource and power efficiency:} Given its aggressive design, Ax-FPM achieves the lowest power consumption and resource utilization.  
%\textbf{(ii) Robustness:} While other approximate designs had a positive impact on CNNs robustness, we noticed that Ax-FPM achieves the highest enhancement. 
%\textbf{(iii) Accuracy:} CNNs are highly approximation-tolerant, and even aggressive multiplier design didn't impact overall accuracy. Ax-FPM, as well as HEAP have insignificant impact on accuracy. 

%While these observations give an insight on the potential generalizability of AC positive impact, Ax-FPM is a non dominated design outperforming other design in both robustness, power consumption and resource utilization, while having the same overall accuracy level. 
%Hence, in the remainder of the paper, we will focus only on Ax-FPM based DA. Further details and results can be found in the Appendix.

\label{sec:AppxCNN}

\section{Can DA help in defending against adversarial attacks? }

In this section, we empirically explore the robustness properties of DA under a number of threat models.  % propose to tackle the relevance of the proposed approximation in enhancing the robustness of CNNs against adversarial attacks. 
We first explore the transferability of adversarial attacks where we evaluate whether attacks crafted for exact CNNs transfer to approximate CNNs. We then consider direct attacks against approximate CNNs in both black and white-box settings.  

%To further study the impact of approximation on CNNs robustness, we comparatively evaluate the sensitivity of approximate CNNs to input gaussian noise.

\subsection{Experimental Setup and Methodology}
In this section, We test two structures: approximate full-precision multiplier (with approximate $24\times 24$ mantissa multiplier) and approximate BFloat16 multiplier (with approximate $8\times 8$ mantissa multiplier).

\subsubsection{Approximate FP32}
The first benchmark we use is the LeNet-5 CNN architecture %\cite{lenet5}
along with the MNIST database \cite{mnist}, which implements a classifier for handwritten digit recognition. The MNIST consists of 60,000 training and 10,000 test images with 10 classes corresponding to digits. Each digit example is represented as a gray-scale image of $28 \times 28$ pixels, where each feature corresponds to a pixel intensity normalized between 0 and 1.
We also use the AlexNet image classification CNN %\cite{Alexnet} 
along with the CIFAR-10 database \cite{CIFAR}.  CIFAR-10 consists of 60,000 images, of dimension $64 \times 64 \times 3$ each and contains ten different classes.  LeNet-5 consists of two convolutional layers, two max-pooling layers, and two fully connected layers.  AlexNet uses five convolution layers, three max-pooling layers, and three fully connected layers. The rectified linear unit (ReLU) was used as the activation function in this evaluation, along with a dropout layer to prevent overfitting.

Our implementations are built using the open source ML framework PyTorch \cite{PyTorch}.
%We use the Adam optimization algorithm to train the LeNet-5 classifier.  For Alexnet, we use Stochastic Gradient Descent (SGD) with a learning rate equal to  $0.01$ and $0.001$, respectively.  
Note that \emph{no retraining} is applied in DA, we rather use the same hyper-parameters obtained from the original (exact) classifier; we simply replace the multipliers with the approximate multiplier.

Our reference exact CNNs are conventional CNNs based on exact convolution layers with the format Conv2d provided by PyTorch.  In contrast, the approximate CNNs emulate the 32-bit Ax-FPM functionality and replace the multiplication in the convolution layers with Ax-FPM in order to assess the behavior of the approximate classifier. %This resulted in very slow simulations, which limited the test cases in terms of input size and classifier architecture. 

Since we are simulating a cross-layer (approximate) implementation from gate-level up to system-level, the experiments (forward and backward gradient) are highly time consuming, which limited our experiments to the two datasets MNIST and CIFAR-10.

%\ihsen{ $\xleftarrow[]{}$ let's keep this to the end (discussion may be)} 
Except for the black-box setting where the attacker trains his own reverse-engineered proxy/substitute model, the approximate and exact classifiers share the same pre-trained parameters and the same architecture; they differ only in the hardware implementation of the multiplier.

\subsubsection{Approximate BFloat16}
% \noindent\colorbox{Yellow}{%
%     \parbox{\dimexpr\linewidth-2\fboxsep}
% {
For the proposed approximate BFloat16 multiplier based experiments. Tests were carried out on an 8th Generation Intel core i7-8750H CPU with 12 cores.  
In order to overcome the issue of extremely slow inference, we used parallel computing, in particular we take advantage of the \textbf{Joblib} \cite{joblib} Python set of tools; using $8$ workers to compute the matrix multiplication per channel in parallel in the convolution layer. %}}

% \noindent\colorbox{Yellow}{%
%     \parbox{\dimexpr\linewidth-2\fboxsep}
% {
%Without the loss of generality 
%\ihsen{@Amira: what do you mean by loss of generality?} \amira{deleted}
We choose three pre-trained models over the MNIST, CIFAR-10, and ImageNet datasets as target models of attacks and defenses. LeNet5 and AlexNet achieve $98.89\%$ and $81\%$ classification accuracy, respectively. VGG-11 model reaches  $90.5\%$ and $76.3\%$ top-5 and top-1 accuracy, respectively. For adversarial example generation, we choose images from the MNIST, CIFAR-10, and ImageNet Validation dataset that are correctly classified by the baseline models: LeNet-5, AlexNet, and VGG-11.
For the evaluation of the defenses, we use FGSM, a commonly used baseline attack. We also used PGD, currently the state-of-the-art attack for $l_\infty$ metric. To ensure a stronger adversary, %we follow recommendations in \cite{recomm} and 
we try achieving maximum confidence rather than minimum distance. This includes making sure that the PGD attack runs through all optimization steps, even if the attack appears to have succeeded on a previous iteration. By default, all experiments will perform PGD with 40 attack iterations.
%}}

\subsection{Do Adversarial Attacks on Exact CNNs Transfer to an Approximate CNN ?}
\label{sec:transExApx}

\begin{figure}[!htp]
\centering
\includegraphics[width=\columnwidth]{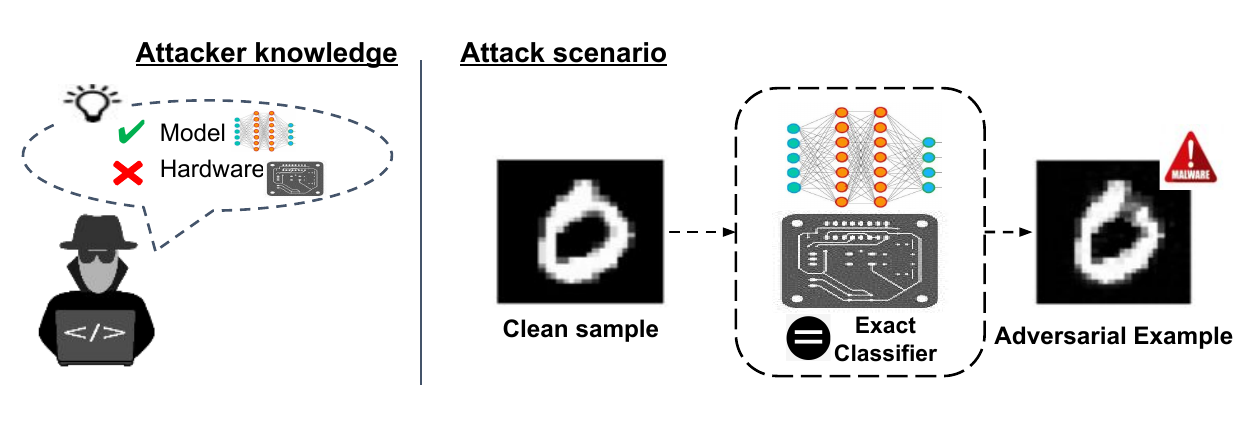}
\caption{Transferrability attack scenario.} %\ihsen{please use "grey" or "gray" (in the figure vs text)}}
\label{grey-box}
\end{figure}

\textbf{Attack Scenario.} In this setting, the attacker has full knowledge of the classifier architecture and hyper-parameters, but without knowing that the model uses approximate hardware.  An example of such scenario could be in case an attacker correctly guesses the used architecture based on its widespread use for a given application (e.g., LeNet-5 in digit recognition), but is unaware of the use of DA as illustrated in Figure~\ref{grey-box}.

\noindent\textbf{Transferability Analysis.} The classifier generates adversarial examples using the set of algorithms in Table~\ref{grey-boxresult} and assume that the exact classifier from Lenet-5 trained on the MNIST dataset. Notice that the hyperparameters, as well as the structure of the network, are the same between the exact and the approximate classifiers.  The attacker then tests the adversarial examples against the approximate classifier.  %The only difference is the underlying approximate hardware for the defender side. %We don't test the entire MNIST dataset due to time and computational constraints as emulating the multiplication to the bit level makes the testing process time consuming and computationally expensive. We choose a subset of the original dataset where different classes are randomly chosen and equivalently represented. %We use the Foolbox \cite{foolbox} tool to generate the adversarial examples.
Table \ref{grey-boxresult} presents the attacks respective success rates. We notice that the DA considerably reduces the transferability of the malicious samples and, by consequence, increases the robustness of the classifier to this attack setting. We observed that the robustness against transferability is very high, and reaches $99\%$ for $C\&W$ attack.

%Please note that this work is the first that explores the robustness properties of CNNs built using approximate circuits.

\begin{table}[!htp]
\small
  \caption{Attacks transferability success rates for MNIST.}
  \label{grey-boxresult}
  \begin{tabular}{ccc}
    \toprule
    \textbf{Attack method} & \textbf{Exact LeNet-5}  & \textbf{Approximate LeNet-5} \\
    \midrule
        FGSM          &   100\%    & 12\%    \\
        PGD           &   100\%    & 28\%    \\
        JSMA          &   100\%    & 9\%   \\
        C\&W          &   100\%    & 1\%    \\
        DF	          &   100\%    & 17\%   \\ 
        LSA	          &   100\%    & 18\%   \\ 
        BA	          &   100\%    & 17\%   \\ 
        HSJ	          &   100\%    & 2\%    \\         
  \bottomrule
\end{tabular}
\end{table}

We repeat the experiment for AlexNet with CIFAR-10 dataset. For the same setting, the success of different adversarial attacks is shown in Table \ref{attackAlexnet}.  While more examples succeed against the approximate classifier, we see that the majority of the attacks do not transfer.  Thus, DA offers built-in robustness against transferability attacks.  %We found that our proposed approach helps enhancing the resiliency of our classifier in consistency with what we proved in the theoretical study.

\begin{table}[!htp]
\small
\centering
  \caption{Attacks transferability success rates for CIFAR-10.}%\nael{No DF results?}
  \label{attackAlexnet}
  \begin{tabular}{ccc}
    \toprule
    \textbf{Attack method} & \textbf{Exact AlexNet}  & \textbf{Approximate AlexNet} \\
    \midrule

        FGSM          &   100\%    & 38\%    \\        
        PGD           &   100\%    & 31\%    \\
        JSMA          &   100\%    & 32\%   \\        
        C\&W          &   100\%    & 17\%    \\
        DF	          &   100\%    & 35\%   \\ 
        LSA	          &   100\%    & 36\%   \\ 
        BA	          &   100\%    & 37\%   \\ 
        HSJ	          &   100\%    & 12\%    \\ 
  \bottomrule
\end{tabular}
\end{table}

Notice that, unlike other state-of-the-art defenses, our defense mechanism protects the network without relying on the attack details or the model specification and without any training beyond that of the original classifier.  
Unlike most of the perturbation-based defenses that degrade the classifier’s accuracy on non-adversarial inputs, our defense strategy significantly improves the classification robustness with no baseline accuracy degradation, as we will show in Section~\ref{CNNaccuracy}.  % (this aspect is detailed in Section \ref{sec:perf_impl}.% negligible impact on non-adversarial images.%, as we showed in the previous subsection. 

%\ihsen{In the remainder of the paper, we will focus only on Ax-FPM based DA. In fact, our design outperforms HEAP-based DA in both robustness (Table \ref{grey-boxresult}) and resource utilization, while having the same overall accuracy level as shown later in Section \ref{CNNaccuracy}.} 

\subsection{Can We Attack an Approximate CNN?}

In the remaining attack models, we assume that the attacker has direct access to the approximate CNN. We consider both black-box and white-box settings.% where a powerful attacker knows the full internal details of the approximate classifier.

\begin{figure}[!htp]
\centering
\includegraphics[width=\columnwidth]{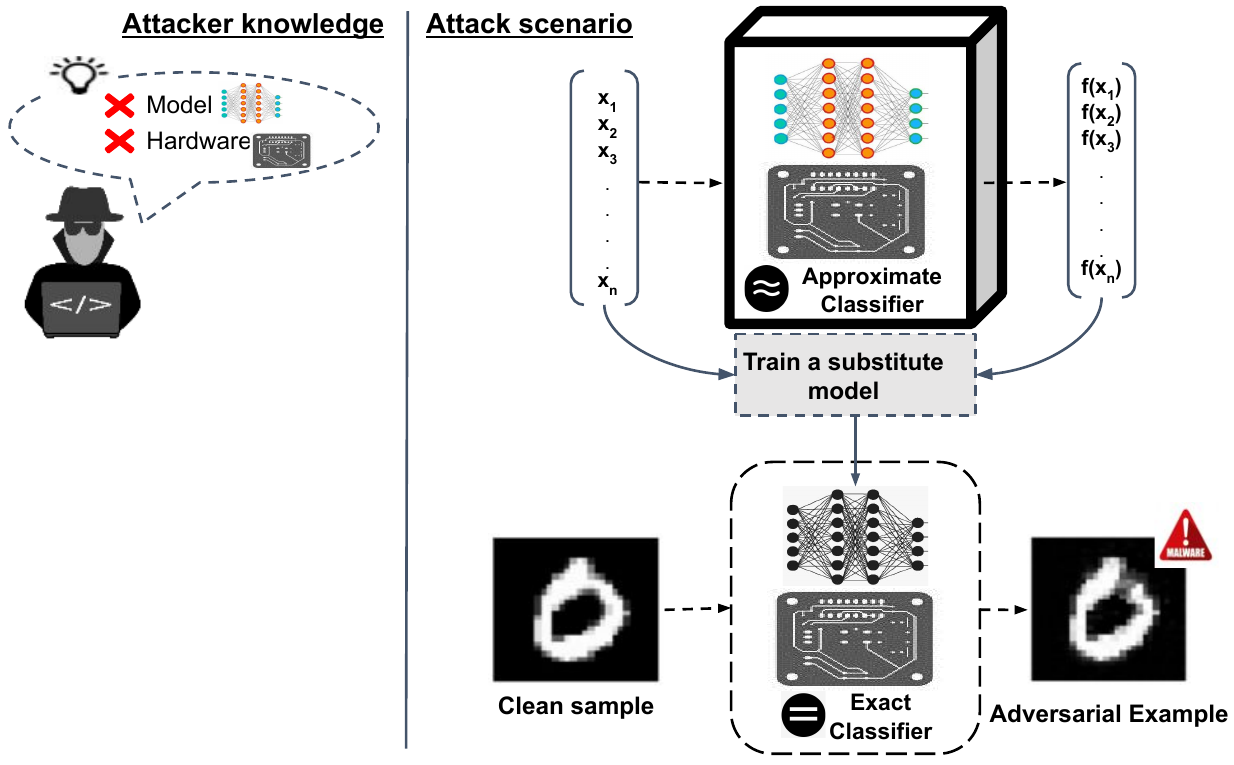}
\caption{Black-box attack scenario. }
\label{black-box}
\end{figure}

\textbf{Black-box Attack.} In a black-box setting, the attacker has no access to the classifier architecture, parameters, and the hardware platform but can query the classifier with any input and obtain its output label.  In a typical black-box attack, the adversary uses the results of many queries to the target model to reverse engineer it.  Specifically, the adversary trains a substitute (or proxy) using the labeled inputs obtained from querying the original model (see Figure \ref{black-box}). % Notice that since the approximate and the exact classifiers have practically the same classification output labels (as we show in Section \ref{CNNaccuracy}), the substitute model is practically the same for the two models under attack. \nael{Is this an important point to make here?}
We also conduct a black box attack on the exact classifier and evaluate how successful the black box attack is in fooling it.  Essentially, we are comparing the black-box transferability of the reverse-engineered models to the original models for both the exact and the approximate CNNs. %Each adversarial example successfully transferred to the exact CNN, its transferability will be evaluated for the proposed approximate CNN.

In Table \ref{black-boxresult}, we present the attack success ratios for the exact CNN and the approximate/DA CNN. DA increases resilience to adversarial attacks across various attacks and for both single-step and iterative ones: it achieves $73\%$ classification success on adversarial examples in the worst case and the defense succeeded in up to $100\%$ of the examples generated by C\&W, PGD, and HSJ respectively.  %ithout re-training the classifier or any additional defense strategy. 
\begin{table}[!htp]
\small
\centering
  \caption{Black-box attacks success rates for MNIST.}
  \label{black-boxresult}
  \begin{tabular}{ccc}
    \toprule
    \textbf{Attack method} & \textbf{Exact LeNet-5}  & \textbf{Approximate LeNet-5} \\
    \midrule
        FGSM          &   100\%    & 22\%    \\
        PGD           &   100\%    & 0\%    \\
        JSMA          &   100\%    & 13\%   \\
        C\&W          &    100\%   & 0\%    \\
        DF	          &   100\%    & 25\%   \\ 
        LSA	          &   100\%    & 26\%   \\ 
        BA	          &   100\%    & 27\%   \\ 
        HSJ	          &   100\%    & 0\%    \\ 
  \bottomrule
\end{tabular}
\end{table}
%\nael{Why is the order in this table different from the order in the original attack table?}
%\subsubsection{White-box Attack }

%\subsection{White-box results}

\begin{figure}[!htp]
\centering
\includegraphics[width=\columnwidth]{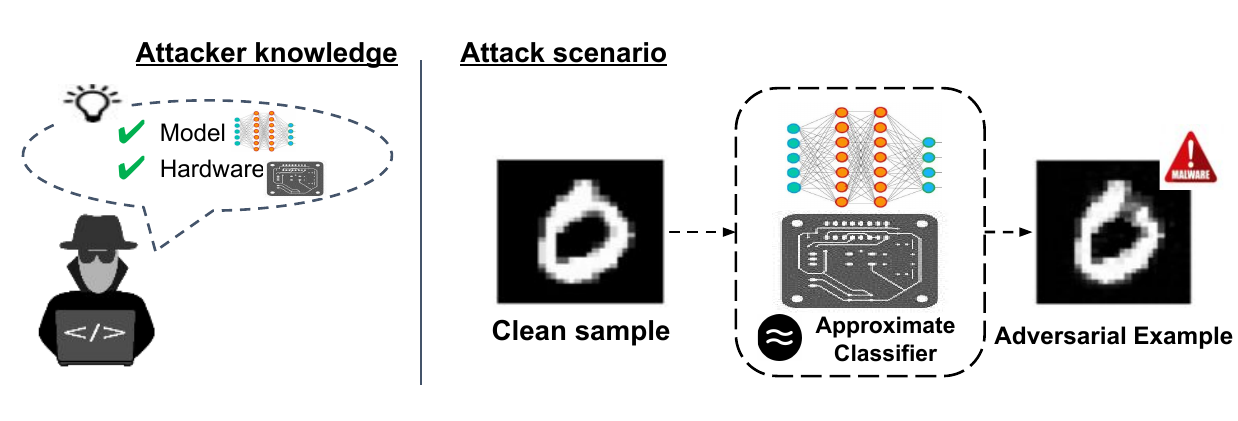}
\caption{White-box attack scenario.}
\label{white-box}
\end{figure}

\textbf{White-box Attack (FP32).} In this setting, the attacker has access to the approximate hardware along with the victim model architecture and parameters, as shown in Figure \ref{white-box}. In particular, the adversary has full knowledge of the defender model, its architecture, the defense mechanism, along with full access to approximate gradients used to build the gradient-based attacks. In essence, the attacker is aware of our defense, and can adapt around it with full knowledge of the model and the hardware, which is a recommended methodology for evaluating new defenses~\cite{carlini_gift}.

In this scenario, we assume a powerful attacker with full access to the approximate classifier's internal model and can query it indefinitely to directly create adversarial attacks.  Although DA in production would normally reduce execution time, in our experiments, we \emph{emulate} the 32-bit Ax-FPM functionality within the approximate classifier.  As a result, this makes inference extremely slow: on average, it takes $5$ to $6$ days to craft one adversarial example on an 8th Gen Intel core i7-8750H processor with NVIDIA GeForce GTX 1050. This led us to limit the white-box experiments; we use only two of the most efficient attacks in our benchmark: $C\&W$ and DeepFool attacks, and for a limited number of examples selected randomly from our test set for different classes. 

In a white box attack, with an unconstrained noise budget, an adversary can always eventually succeed in causing an image to misclassify. Thus, robustness against this type of attack occurs through the magnitude of the adversarial noise to be added: if this magnitude is high, this may exceed the ability of the attacker to interfere, or cause the attack to be easily detectable. %in two ways: (1) The magnitude of the adversarial noise to be added: if this magnitude is high, this may exceed the ability of the attacker to interfere, or cause the attack to be easily detectable; and (2) the number of iterations, and consequently the time for producing adversarial examples: an attack that exceeds a certain limit of queries might be detected and stopped, such as the case of Cloud-based machine-learning a service platforms.   %\nael{Did we measure this second one? Could be good to see this.} \amira{No sorry we don't have this info}
%In an attempt to attack the approximate classifier in a white-box setting, we reduce the attacks benchmark  DeepFool attack. %With a high-dimensional and non-linear image space, finding the required perturbation to push the resulting image towards the decision boundary is non-trivial. 
%This attack uses a linearized approximation of the boundary to iteratively solve the problem. In each iteration, a step is made, i.e. a perturbation is added to the image, in the direction of the linearized boundary until the image crosses the original decision boundary.
%The limited number of experiments was due to, as stated previously, time and computational constraints. In this scenario, we assume a powerful attacker with full access to the approximate classifier's internal model and can query it indefinitely to directly create adversarial attacks
\begin{figure*}[!t]
    \begin{subfigure}{.5\textwidth}
        \centering
        \includegraphics[width=0.95\columnwidth]{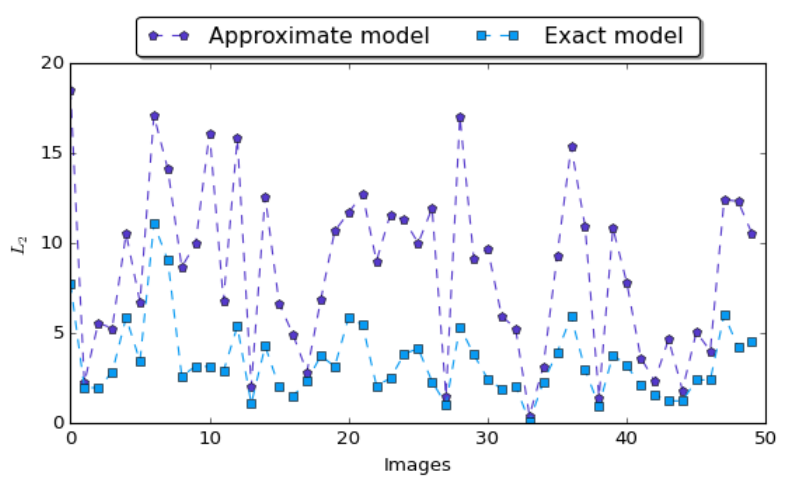}
        \caption{ DeepFool  }
        \label{L2}
    \end{subfigure}
    \begin{subfigure}{.5\textwidth}
        \centering
        \includegraphics[width=0.95\columnwidth]{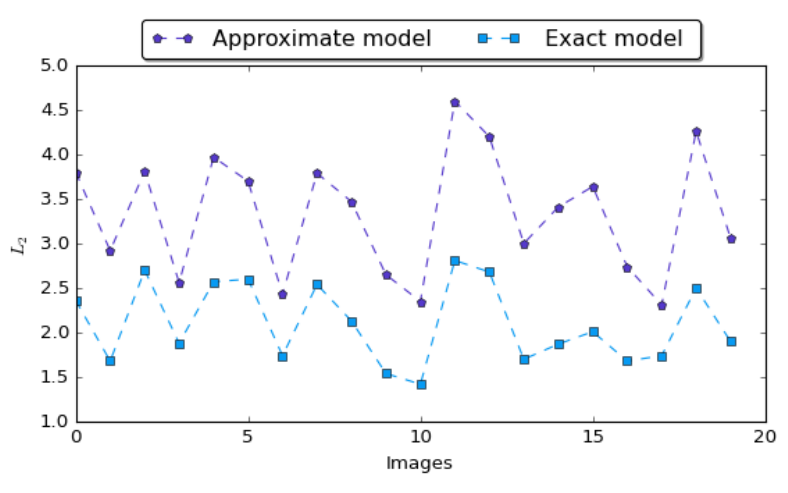}
        \caption{ C\&W  }
        \label{L2_C&W}
    \end{subfigure}
\caption{Required $L_2$ perturbations by C\&W and Deepfool attacks, to generate MNIST adversarial examples.}
\end{figure*}

Figures \ref{L2} and \ref{L2_C&W}, respectively, present different measures of $L_2$ for adversarial examples crafted using DF and $C\&W$ attacking both a conventional CNN and an approximate CNN. We notice that the distance between a clean image and the adversarial example generated under DA is much larger than the distance between a clean sample and the adversarial example generated for the exact classifier. On average, a difference of $5.12$ for $L_2$-DeepFool attacks and $1.23$ for $L_2$-C\&W attack. %Herex\nael{Amira, does is it a good idea to also show the increase in L2 (e.g., saying something like "approximately doubling the L2 distance needed for the attack to succeed")}
%\nael{How much on average?  Can we help make the readers understand if this is significant?} \amira{done}  
This observation confirms that DA is more robust to adversarial perturbations since the magnitude of the adversarial noise has to be significantly higher for DF to fool DA successfully.  

%To understand the implication of this higher robustness in terms of observable effects on the input image,  we also show the Peak Signal to Noise Ratio (PSNR) and the Mean Square Error (MSE) in Figures \ref{MSE&PSNR_DF} and \ref{MSE&PSNR}; these are two common measures of the error introduced in a reconstruction of an image. Specifically,  MSE represents the average of the squares of the "errors" between the clean image and the adversarial image. The error is the amount by which the values of the original image differ from the distorted image. PSNR is an expression for the ratio between the maximum possible value (power) of a signal and the power of distorting noise that affects the quality of its representation. It is given by the following equation: $P S N R=20 \log _{10}\left(\frac{M A X_{x}}{\sqrt{M S E}}\right)$. The lower the PSNR, the higher the image quality degradation is.

%We notice that the adversarial examples generated for DA is more noisy than adversarial examples generated for an exact classifier. The PSNR difference reaches $4 dB$ for C\&W and $7.8dB$ for DeepFool. Moreover, on average, the DA-dedicated adversarial examples have $6$ times, and $3$ times more MSE than the exact classifier-dedicated adversarial examples for $C\&W$ and DeepFool attacks, respectively. 

% \begin{figure}[!htp]
% \centering
% \includegraphics[width=0.9\columnwidth]{figures/L2_C&W.pdf}
% \caption{Required perturbations by C\&W attack, measured using $L_2$ distance, to generate MNIST adversarial examples for approximate and exact classifiers. }
% \label{L2_C&W}
% \end{figure}

\begin{figure}[!htp]
\centering
\includegraphics[width=\columnwidth]{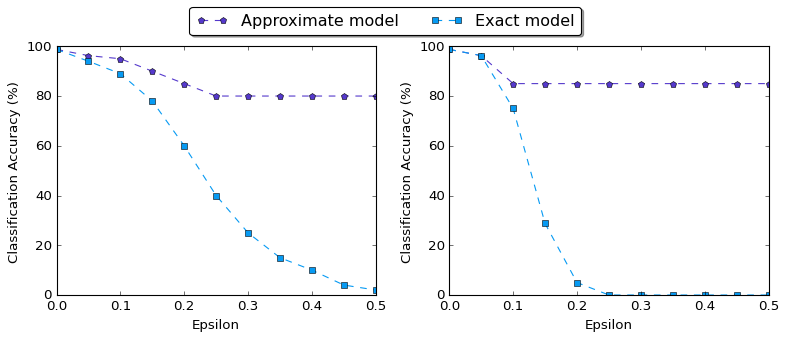}
\caption{Model accuracy for different noise budget under white-box attack on MNIST using (left) FGSM and (right) PGD.}
\label{fig_pgd_mnist}
\end{figure}

\label{white_box_bf}
% \noindent\colorbox{Yellow}{%
%     \parbox{\dimexpr\linewidth-2\fboxsep}
% {
\textbf{White-box Attack (BF16).} In this setting, we test models built using the proposed approximate BFloat16 multiplier.
%}}
% \noindent\colorbox{Yellow}{%
%     \parbox{\dimexpr\linewidth-2\fboxsep}
%{
\noindent \textbf{\textit{MNIST.}} In this setting, we assume a powerful adversary with full access to the defense mechanism and the victim model architecture and parameters. To evaluate the effectiveness of our proposed method, we measure the model's classification accuracy of adversarial images constructed using different attacks.
In Figure \ref{fig_pgd_mnist}, we report the classification accuracy under $l_\infty$ FGSM and PGD attacks for different perturbation magnitudes when testing images from MNIST dataset. We notice that even for high amounts of noise, our defended model maintained high resilience. For instance, using our defense, the model's accuracy remains up to $80\%$ and $85\%$ under FGSM and PGD, respectively, with a noise budget of $\varepsilon = 0.5$. 
The images that are consistently correctly classified regardless of the noise budget correspond to the cases where we have a vanishing gradient, i.e, an input gradient almost equal to zero. This vanishing gradient prevents the iterative noise update from evolving in the direction of maximizing the loss function. Further details can be found in Section \ref{impact_grad}.
%\ihsen{we might point to Section 6.2. for further details about the gradient investigation.} \amira{done}
%}} 

\begin{figure*}[!t]
	\begin{subfigure}{.5\textwidth}
		\centering
        \includegraphics[width=0.85\linewidth]{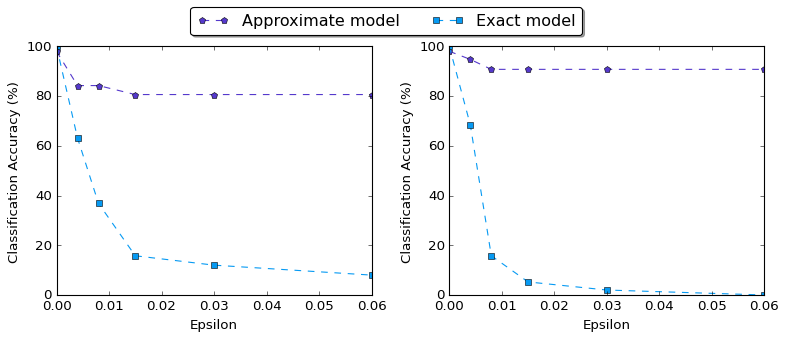}
        \caption{CIFAR-10 using (left) FGSM and (right) PGD.}
        \label{fig_pgd_cifar}
    \end{subfigure}
    \begin{subfigure}{.5\textwidth}
		\centering
        \includegraphics[width=0.85\linewidth]{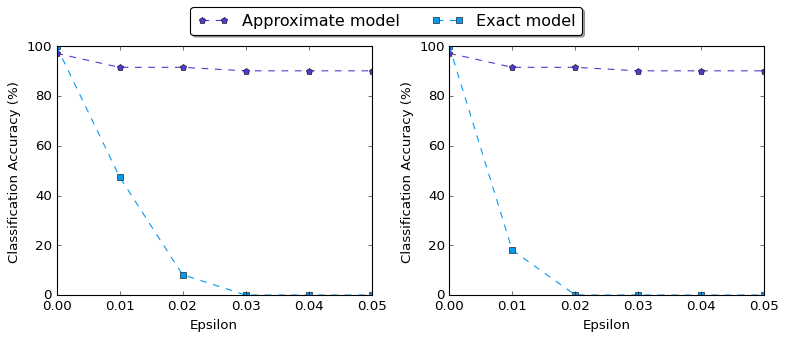}
        \caption{ImageNet using (left) FGSM and (right) PGD.}
        \label{fig_pgd_imagenet}
    \end{subfigure}
    
	\caption{Model accuracy for different noise budgets under white-box attack.}
	\label{fig:cifar_Imgnet}

\end{figure*}

% \noindent\colorbox{Yellow}{%
%     \parbox{\dimexpr\linewidth-2\fboxsep}
% {
\noindent \textbf{\textit{CIFAR-10.}} Figure\ref{fig_pgd_cifar} reports the classification accuracy of images from CIFAR-10 dataset when attacked using FGSM and PGD. We notice similar trends to previous results. For example, DA enhances the robustness of the model to reach $80\%$ and $90\%$ under FGSM and PGD attack for a noise budget of $0.06$.
% }}

% \noindent\colorbox{Yellow}{%
%     \parbox{\dimexpr\linewidth-2\fboxsep}
% {
\noindent \textbf{\textit{ImageNet.}}
Figure \ref{fig_pgd_imagenet} summarizes the effectiveness of our defense against FGSM and PGD attacks. Similar to previous experiments, the approximate hardware prevent the attacker from generating efficient AE for deeper networks and more complex data. Even with a high amount of injected noise ($\varepsilon = 0.06$), our model accuracy still reaches $90\%$ under PGD attack. 
%}}
%\begin{figure}[!t]
%\centering
%\includegraphics[width=3.5in]{figures/class_accuracy_FGSM_imagent.PNG}
%\caption{Model accuracy for different noise budget under white-box attack on ImageNet using FGSM.}
%\label{fig_fgsm_imagenet}
%\end{figure}

%\begin{figure}[!t]
%\centering
%\includegraphics[width=3.5in]{figures/class_accuracy_PGD_imagent.PNG}
%\caption{Model accuracy for different noise budget under white-box attack on ImageNet using PGD.}
%\label{fig_pgd_imagenet}
%\end{figure}

\begin{figure}[!htp]
\centering
\includegraphics[width=3in]{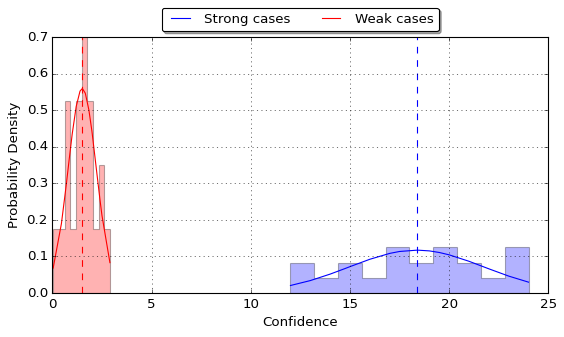}
\caption{Gaussian distribution of Confidence (pre-softmax) for strong vs weak cases.}
\label{fig_conf_dist}
\end{figure}

% \begin{figure}[!htp]
% \centering
% \includegraphics[width=\columnwidth]{figures/imagenet_bf16.png}
% \caption{Model accuracy for different noise budget under white-box attack on ImageNet using (left) FGSM and (right) PGD.}
% \label{fig_pgd_imagenet}
% \end{figure}
% \noindent\colorbox{Yellow}{%
%     \parbox{\dimexpr\linewidth-2\fboxsep}
% {
Higher classification accuracy can be noticed for small noise magnitude. %These cases correspond to images where the model does not classify this image with high confidence. 
%\ihsen{Not sure if I understood this correctly, neither am I sure of its importance here-- } \amira{deleted}
As illustrated in Figure \ref{fig_conf_dist}, the cases that we have named the strong cases correspond to image samples classified with high confidence and are robust to a high noise magnitude, whereas what we have named weak cases correspond to image samples classified with much lower confidence and are robust to very low noise amplitude but can be eventually transformed into adversarial examples. We believe that these cases are the reason behind the higher classification accuracy for small noise magnitude.
%}}

We can conclude that DA provides substantial built-in robustness for all three attack models we considered.  Attacks generated against an exact model do not transfer successfully to DA.  Black-box attacks also achieve a low success rate against DA.  Finally, even white-box attacks require substantial increases in the injected noise to fool DA for $L_2$-based attacks and are inefficient even under high noise magnitude for $L_\infty$-based attacks. Next, we probe deeper into DA's internal behavior to provide some intuition and explanation for these observed robustness advantages.

%\begin{figure}[!htp]
%\centering
%\includegraphics[width=\columnwidth]{figures/MSE_PSNR_DF.png}
%\caption{MSE and PSNR values for the generated adversarial examples using DeepFool method when attacking the approximate and the exact classifiers.}
%\label{MSE&PSNR_DF}
%\end{figure}

%\begin{figure}[!htp]
%\centering
%\includegraphics[width=\columnwidth]{figures/MSE_PSNR.png}
%\caption{MSE and PSNR values for the generated adversarial examples using $L_2$-C\&W method when attacking the approximate and the exact classifiers.}
%\label{MSE&PSNR}
%\end{figure}

\label{sec:secu}

\section{How does DA help CNN robustness?}
In this section, we probe into the DA classifier's operation to attempt to explain the robustness advantages we observed empirically in the previous section.  %Given the promising empirical results shown in the previous section, we attempt in this section to provide an analysis of the defensive approximation mechanism. %the eventual mechanisms that are involved in the security enhancement due to approximate computing. 
While the explainability of deep neural networks models is a known hard problem, especially under adversarial settings~\cite{samek2019explainable}, we attempt to provide an overview of the mechanisms that we think are behind the DA impact on security. %Before proceeding further, notice that our purpose in this section is not to provide a definitive argument about the explanation of the AC impact on robustness, but rather we view it as an initial step towards drawing an explanation that connects the circuit-level approximation to the model robustness for future work to build upon.  

\subsection{Impact of approximation on model confidence}
We study the impact of the approximation on CNNs' confidence and generalization property. We follow this analysis in the Appendix with a mathematical argument explaining the observed robustness based on recent formulations by Lecuyer et al.~\cite{snP2019_certif}.

The output of the CNN is computed using the \emph{softmax} function, which normalizes the outputs from the fully connected layer into a likelihood value for each output class.  Specifically, this function takes an input vector and returns a non-negative probability distribution vector of the same dimension corresponding to the output classes.   In this section, we examine the impact of approximation on the observed classifier confidence. We compare the output scores of an exact and an approximate classifier for a set of $1000$ representative samples selected from the MNIST dataset: $100$ randomly selected from each class. We define the classification confidence, $C$, as the difference between the true class $l$'s score and the "runner-up" class score, i.e., the class with the second-highest score. $C$ is expressed by $C = output[l] - max_{j \neq l} \{output[j]\}$.  The confidence ranges from $0$ when the classifier gives equal likelihood to the top two or more classes, to $1$ when the top class has a likelihood of $1$, and all other classes $0$.  

\begin{figure}[!htp]
\centering
\includegraphics[width=0.9\columnwidth]{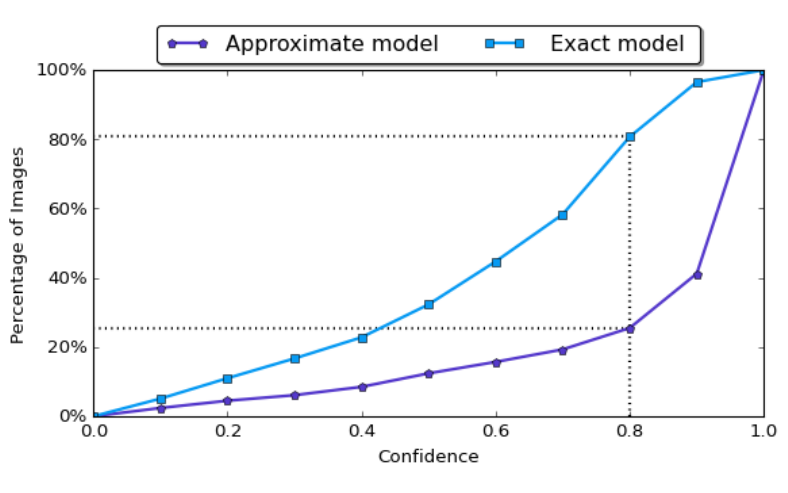}
\caption{Cumulative distribution of confidence.}
\label{conf_dis}
\end{figure}

We plot the cumulative distribution of confidence for both classifiers in Figure \ref{conf_dis}.  DA images have higher confidence; for example, in images classified by the exact classifier, less than $20\%$  had higher than $0.8$ confidence.   
On the other hand, for the approximate classifier, $74.5\%$ of the images reached that threshold.

Compared to the baseline feature maps generated from an exact convolution operation, for the same pre-trained weights, our approximate convolution highlights further the features. Recall that the multiplier injected noise is higher when input numbers are higher (i.e., there is a high similarity between the kernel and the input data) and lower when the inputs are lower (when the similarity is small), as shown in Figure \ref{fig_sim}. We believe that these enhanced features continue to propagate through the model resulting in a higher probability for the predicted class. This higher confidence requires thereby higher noise to decrease the true label's likelihood, and increase another label's.  

\subsection{Impact of approximation on model gradient}
\label{impact_grad}
% \noindent\colorbox{Yellow}{%
%     \parbox{\dimexpr\linewidth-2\fboxsep}
% {
\begin{figure*}[!htp]
\centering
\includegraphics[width=.8\textwidth,height=2.3in]{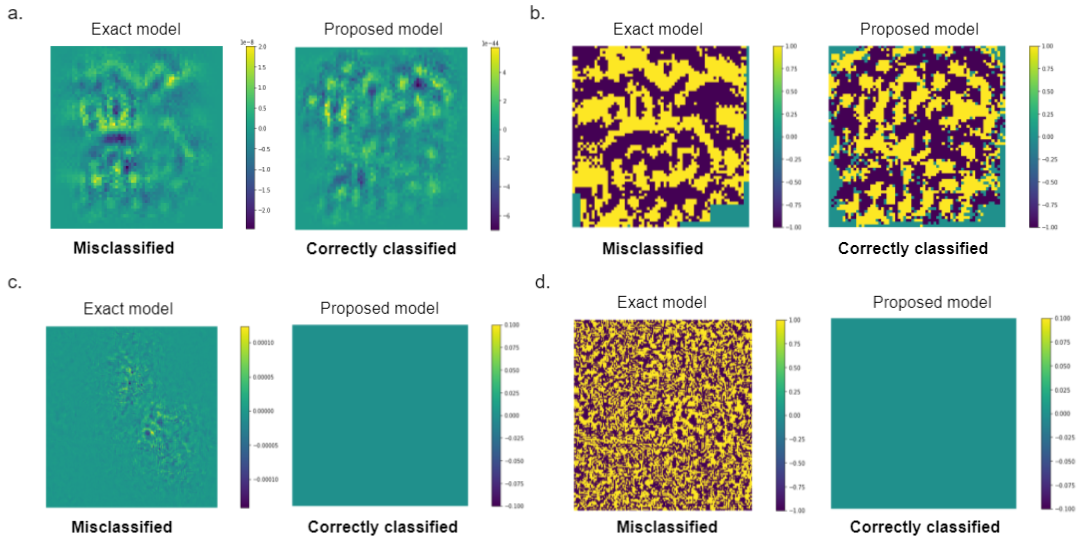}
\caption{a. \& c. Input gradient and b. \& d. Sign of Input gradient for different models (top) for CIFAR-10 dataset (bottom) for ImageNet dataset.}
\label{fig_grad_input_sign_imagenet}
\end{figure*}

In this subsection, we analyze the approximate computing impact on the gradient, which is a key element in adversarial attacks generation process. 

In this experiment, we perform the PGD attack and visualize the input gradient used to update the adversarial noise as it represents the direction, in the input space, that results in the larges change in loss for different models with different combinations of approximate convolution and FC layers.

%\ihsen{@Amira: let's beef-up this section; please explain what you did, conceptually and implementation-wise. }
We notice that stacking approximate layers results in vanishing input gradient as shown in Figure \ref{fig_grad_input_sign_imagenet}.c or makes almost equal to zero a significant number of input gradient elements as in Figure \ref{fig_grad_input_sign_imagenet}.a. In order to generate an adversarial example, the sign of the input gradient, will be multiplied by the noise $\varepsilon$, as in Figures \ref{fig_grad_input_sign_imagenet}.b and \ref{fig_grad_input_sign_imagenet}.d. A nullified gradient means no $\varepsilon$ step is performed in that case the adversary is unable to create an efficient AE when using an approximate model. 
%}}

\label{sec:how}

\section{How does DA compare to other reduced precision techniques?}

In this section, we investigate the impact of other reduced precision techniques on robustness. We first compare DA to DQ, and then study the impact of using BFloat16 data representation on the system performance and robustness.

%\label{sec:quan}
\subsection{Defensive Quantization}
%Defensive quantization \cite{lin2019defensive} 

Defensive quantization \cite{DQ} was proposed to jointly optimize the efficiency and robustness of deep learning models. A 4-bit quantized model was trained on CIFAR-10 using the Dorefa-Net method \cite{zhou2018dorefanet}. %The model architecture is detailed in Appendix \ref{mod_archi}. 
We consider two ways of quantization: (i) Weight quantization, where only the weights are quantized, and (ii) Full quantization where the weights %\ihsen{CNN input? no?} 
of each convolutional and dense block and the output of each activation function are quantized. %%%%%%%%%%
In Table \ref{quan_result4}, we report transferability between exact (32-bit floating point), approximate model (using DA), fully quantized and weight-only quantized model.
We notice that DA is almost  two times more robust against transferability attacks than DQ under FGSM, PGD and C\&W attacks. We believe that this is due to the difference in terms of noise distribution between DA and DQ. In fact, while DQ-induced noise tends to make the initial decision boundary smoother, the input-dependent noise in DA makes its decision boundary randomly different from the initial model.

\begin{table}[!htp]
\small
\centering
  \caption{Comparing attacks transferability success rates for CIFAR-10 when using DA and DQ.}
  \label{quan_result4}
  \begin{tabular}{ccccc}
    \toprule
                     &          & \textbf{DA:} & \textbf{DQ:} & \textbf{DQ:} \\
    \textbf{Attack method} & \textbf{Exact}  & \textbf{FP} & \textbf{Full } & \textbf{Weight-only } \\
    \midrule
        FGSM          &   100\%    & 38\%  & 60\%  & 61\% \\
        PGD           &   100\%    & 31\%  & 74\% & 73\%\\
        C\&W          &   100\%    & 17\%  & 68\% &  68\% \\
        
  \bottomrule
\end{tabular}
\end{table}

\subsection{Conventional BFloat16}

To evaluate the impact of BFloat16 on deep neural networks robustness, we use Pytorch framework \cite{PyTorch} to implement BFloat16-based CNN architectures and test them for MNIST and CIFAR-10 benchmarks.
We notice that using BFloat16 achieves the same accuracy as the 32-bit floating point. No remarkable change was noticed at the output of the convolutional layers nor in the models confidence. Also as shown in Table \ref{quan_result41}, the BFloat16 model is as vulnerable as the exact model. We believe this is due to the nature of the noise introduced by reducing data representation accuracy. A more detailed discussion can be found in Section \ref{discussion}. 

\begin{table}[!htp]
\small
\centering
  \caption{Comparing attacks transferability success rates for CIFAR-10 when using DA and BFloat16.}
  \label{quan_result41}
  \begin{tabular}{cccc}
    \toprule
    \textbf{Attack method} & \textbf{Exact}  & \textbf{DA} & \textbf{BFloat16 }  \\
    \midrule
        FGSM          &   100\%    & 38\%  & 100\%   \\
        PGD           &   100\%    & 31\%  & 100\% \\
        C\&W          &   100\%    & 17\%  & 100\% \\
        
  \bottomrule
\end{tabular}
\end{table}

\begin{figure}[!htp]
\centering
\includegraphics[width=0.8\columnwidth]{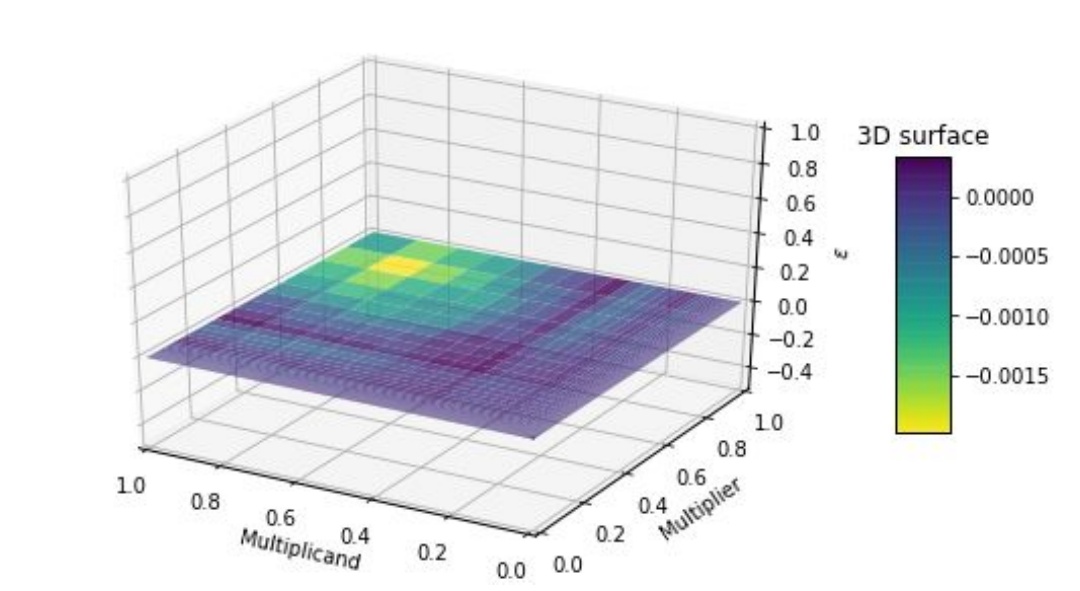}
\caption{Noise introduced by multiplying Bfloat16 while the operands $\in [0,1]$.
}
\label{bfloat16}
\end{figure}

Figure \ref{bfloat16} shows the multiplication noise distribution for $10^8$ randomly generated BFloat16 numbers compared to their corresponding 32-bit floating point. We notice that, in contrast with our approximate multiplier (Figure \ref{appx_mult}), Bfloat16 multiplication results in mostly negative noise with orders of magnitude lower than DA-induced noise. Moreover, the Bfloat16 noise has no specific impact on the model confidence. Accordingly, no improvement in Bfloat16 models robustness was noticed under FGSM, PGD and $C\&W$. 

\label{sec:quan}

\section{Baseline Performance Implications}
\subsection{Impact on Model Accuracy} %(a) same data distribution (b) noise }
\label{CNNaccuracy}

%It is important for a defense mechanism that aims to enhance robustness against adversarial attacks to keep at least an acceptable performance level for clean inputs. In fact, considerably reducing the baseline accuracy, or creating an exploding or vanishing gradient impact that makes the model sensitive to other types of noise undermines the model reliability.  
%In our proposed approach, we maintain the same level of recognition rate even with the approximate noise in the calculations. Counter-intuitively, this data-dependent noise helps to better highlight the input's important features used in the recognition and does not affect the classification process. A drop of $0.01 \%$ in the recognition rate for the case of LeNet-5 and $1\%$ for AlexNet is recorded as mentioned in Table \ref{tab3}. 

%\begin{table}[!htp]
%\small
%\centering
%  \caption{Accuracy results of Float32, approximate model, fully quantized, weight-only quantized and Bfloat16 models.}
%  \label{tab3}
%  \begin{tabular}{ccc}
%    \toprule
%    \textbf{Used Multiplier }& \textbf{MNIST} & \textbf{CIFAR-10}\\
%    \midrule
%        Float32                 & 97.93\%  &   81\%   \\
%        Approximate (DA)	    & 97.67\%  &   80\%     \\ 
%        Fully quantized         & -     &   80\%  \\
%        Weight-only quantized   & -     &  80\% \\
%        Bfloat16                & 97.93  &  81\% \\
%  \bottomrule
%\end{tabular}
%\end{table}

% \noindent\colorbox{Yellow}{%
%     \parbox{\dimexpr\linewidth-2\fboxsep}
% {

When fully approximating (both convolution and FC layers) deeper models, we recorded a drop in models' classification accuracy. In fact, we notice $14\%$ and $26\%$ drop in accuracy for the AlexNet and VGG-11, respectively.
In Table \ref{tab:ML}, we perform an ablation study to track the impact of each approximate layer on deep networks accuracy. We found that the first convolution layer has the highest impact on models accuracy. Specifically, when using an exact first convolution layer, models recover $12\%$ and $23\%$ of their classification accuracy. %\ihsen{should we mention that the security experiments where performed based on this observation, i.e., the first conv was exact all the time? Or you probably mention it and I didn't pay attention?} \amira{done}
Based on this observation, we used an exact first convolution layer when running the security experiments of the approximate BFloat16 multiplier under white-box setting for both CIFAR-10 and Imagenet datasets.

%}}

\begin{table}[!htp]
\centering
\caption{ Impact of approximation on model classification accuracy for a set of clean Inputs from Cifar-10 and ImageNet.}
\begin{tabular}{ccc}
\toprule
                & \multicolumn{2}{c}{\textbf{Top 1 accuracy}}      \\
\textbf{Model}  &  \textbf{CIFAR-10} & \textbf{ImageNet}  \\
\midrule
Full exact model        & 100\%  & 100\% \\
Full approximate model	& 85.7\%  & 73.23\% \\
Exact: 1st conv layer	& 98.34\%  & 97.18\%  \\
Exact: 2nd conv layer	& 93.4\%  & 83.60\%  \\
Exact: 3rd conv layer    & 93.4\%   & 83.60\% \\
Exact: 1st FC layer      & 88.04\%  & 75.4\%  \\
Exact: 2nd FC layer      & 88.04\%  & 75.4\%  \\
Exact: 3rd FC layer      & 88.04\%  & 75.4\%   \\
Exact: all FC layers     & 95\%  & 78.87\%   \\

\bottomrule 
\end{tabular}
\label{tab:ML}
\end{table}
% \noindent\colorbox{Yellow}{%
%     \parbox{\dimexpr\linewidth-2\fboxsep}
% {
A defense mechanism that aims to improve robustness against adversarial attacks must maintain at least a reasonable performance for clean inputs.
For our defense, even with the approximate noise injected in the calculations, approximate models achieve comparable recognition rates to the exact models. %This data-dependent noise helps to better illustrate the important features of the input while having no effect on the classification process.
A drop of less than $1\%$ in the recognition rate for MNIST, less than $2\%$ for CIFAR-10, and $3\%$ for Imagenet is recorded as mentioned in Table \ref{tab3}. 
%}}

\begin{table}[!htp]
\small
\centering
  \caption{Top-1 Accuracy results of eaxct Float32, approximate Float32, exact BFloat16 and approximate Bfloat16 models.}
  \label{tab3}
  \begin{tabular}{cccc}
    \toprule
    \textbf{Used Multiplier }& \textbf{MNIST} & \textbf{CIFAR-10} & \textbf{ImageNet} \\
    \midrule
        Exact Float32           &   98.89\%   &   81\%   &  76.3\%  \\
        Approximate Float32	    &    98.76\%  &   79.34\%   &  73\%  \\ 
        Exact BFloat16          &   98.89\%   &   81\%   & 76.3\%    \\
        Approximate BFloat16    &    98.76\%  &   79.34\%   & 73\%  \\
        
  \bottomrule
\end{tabular}
\end{table}

\subsection{Impact on Performance and Energy Consumption}

% %\begin{figure}[!tp]
% %\centering
% %\includegraphics[width=0.85\columnwidth]{asplos21-templates/figures/delay_energy.pdf}
% %,height=5cm
% %\caption{Energy, delay and EDP of AMA5-based $24\times24$ approximate multiplier normalized to a conventional multiplier.}
% %\label{fig:energy}
% %\end{figure}

% \sout{

Here we show the additional benefit of using AC, especially in the context of power-limited devices, such as mobile, embedded, and Edge devices.
The experiments evaluate normalized energy and delay achieved by the proposed approximate multiplier compared to a conventional baseline multiplier. Multipliers are implemented using $45~ n m$ technology via the Predictive Technology Model (PTM) using the Keysight Advanced Design System (ADS) simulation platform~\cite{ptm}.  

 \begin{table}[!htp]
 \centering
   \caption{Energy and delay Comparison.}
   \label{tab:energy}
   \begin{tabular}{ccc}
     \toprule
     \textbf{Multiplier }& \textbf{Average energy}	& \textbf{Average delay}  	 \\
     \midrule
         Exact Float32        & 1 & 1    \\
         %HEAP                   & 0.49  & 0.46\\
         %Ax-FPM	             & 0.501 & 0.328\\ 
         Approximate Float32                 & 0.487 & 0.29\\ 
         Exact BFloat16          & 0.4   & 0.4   \\
         Approximate BFloat16    &  0.19   &  0.12\\
   \bottomrule
 \end{tabular}
 \end{table}
 
% \sout{
% \noindent\colorbox{Yellow}{%
%     \parbox{\dimexpr\linewidth-2\fboxsep}
% {
Table \ref{tab:energy} compares the energy and delay for the approximate Float32 multiplier, the BFloat16 multiplier, and the approximate BFloat16 normalized to a conventional Float32 multiplier. 
%The used Bfloat16 multiplier has similar architecture to that of Ax-FPM. However, for the mantissa multiplier, we use the conventional Booth multiplier instead of the array multiplier.
Ax-FPM achieves up to $51\%$ and $71$\% savings in energy and delay, respectively, compared to the baseline multiplier. An exact Bfloat16 implementation results in around $60\%$ lower power and delay.  
This comes with no robustness advantages, as mentioned earlier. However, the proposed approximate BFloat16 multiplier ensures a gain in energy of up to $81\%$ and $88\%$ saving in delay along with robustness advantages.
%\ihsen{didn't understand, BFloat and AxFPM have comparable energy gains?}. \amira{fixed}
%}}
 %60\% of exact mantissa multiplier total consumption, adding the power gained by eliminating the rounding unit, the final power saving of the whole FPM exceeds 67\%.
 %The approximate circuit also achieves more than $60\%$ reduction in delay.  AMA5 uses only eight transistors rather than 28; corresponds to 72\% area savings. %
Unlike most of the state-of-the-art defense strategies that lead to power, resource, or timing overhead, DA results in saving energy and area.

%%%%%%%%%%%%%%%%%%%%%%%%%%%%%%%%
%\section{Hardware acceleration}
% \noindent\colorbox{Yellow}{%
%     \parbox{\dimexpr\linewidth-2\fboxsep}
% {
We report in Table \ref{joblib} the obtained speedup when using parallel computing in the inference of different approximate BFloat16-based models. In fact, for samples from ImageNet, we were able to make the model inference almost $10\times$ faster.
%}} % \ihsen{I wonder if we shouldn't value this part somewhere separately-- but definitely its place is not in the security evaluation section. Also, Amira, we might want to explain it in a more conceptual way if possible before going to details and tools (anything interesting in the algorithms for example ?) }

\begin{table}[!htp]
\small
\centering
  \caption{Inference time with and without parallel computing of approximate BFloat16 models.}
  \label{joblib}
  \begin{tabular}{cccc}
    \toprule
      Run-time (s)   &  \textbf{LeNet-5}  & \textbf{AlexNet} & \textbf{VGG-11} \\
    \midrule
    without parallelism &   57    &     1 290    &  69 110\\
    with parallelism    &   9     &     450     & 7 200 \\        
  \bottomrule
\end{tabular}
\end{table}
\label{sec:perf_impl}

\section{Discussion}

This work tackles the problem of robustness to adversarial attacks from a new perspective: approximation in the underlying hardware. DA exploits the inherent fault tolerance of deep learning systems %\cite{iccd18} 
to provide resilience while also obtaining the by-product gains of AC in terms of energy and resources. Our empirical study shows promising results in terms of robustness across a wide range of attack scenarios. We notice that AC-induced perturbations tend to help the classifier generalize and enhances its confidence.  We believe that this observation is possibly due to the specific AC multiplier we used where the introduced noise is input-dependent and non-uniform.
When we observe the effect on the convolution layer, we see higher absolute values when the inputs are similar to the convolution filter. This observation at the feature map propagates through the model and results in enhanced classification confidence, i.e., the difference between the $1^{st}$  and  runner-up classes. 
This aspect of confidence enhancement resembles the smoothing effect observed by some recently proposed randomization techniques~\cite{snP2019_certif}.  

 We believe that our study makes important contributions in demonstrating the general potential of approximate computing in this new dimension.  However, we believe that substantial further research remains which we hope to tackle in our future work:  (1) More work is needed to carefully understand the relationship between the patterns of induced noise and the observed robustness to adversarial attacks to guide the selection of approximation approaches; (2) We would also like to explore whether there is additional protection that results from adapting the approximation function over time; (3) We believe that DA is orthogonal to some of the other AML defenses and, deployed together, they may result in even higher protection against AML; (4) Some AML defenses unintentionally make the model more susceptible to privacy related attacks~\cite{rezaShokry}; we believe that DA does not have a similar effect and would like to study its implication on privacy-preserving in the future; and (5) We would like to explore DA in the context of other learning structures beyond CNNs.
 
 %While we do not claim a full and definitive explanation of this mechanism, we believe that the circuit-level approximate mantissa multiplier has a positive effect on the features highlighting. 

%We think that this work opens a new research direction in tackling the deep learning security problem and encourages researchers to investigate further the use of AC as a potential defense mechanism against adversarial attacks. Besides, we believe that the AC-induced noise can also be useful in a privacy preserving context.%, especially in the ca attacks such as membership inference attacks. 

DA has two important advantages compared to DQ: (1) DA results in input-dependent noise, while DQ results in a deterministic network that can be efficiently reverse engineered and undermined by adaptive white-box attacks; (2) DQ requires retraining/fine-tuning the model to avoid drastic accuracy drop, while DA does not require retraining.

Compared to DA, when proceeding to BFloat16 quantization, we did not notice any improvement in the model robustness, which could be explained by the fact that BFloat16 results in uniform low-amplitude noise distribution that is not sufficient to impact CNNs behavior.

While  we considered full precision floating point CNNs, we believe that DA can also apply to  quantized and sparse networks~\cite{quant,survey_121,NIPS2018_sparce} with similar impact on security. Prior work~\cite{D&T} shows that quantized networks tolerate errors, implying that DA can potentially be deployed without degrading accuracy.  %%Furthermore, we believe that the robustness enhancement introduced by hardware-supported approximation is orthogonal to some other existing robustness approaches (e.g., input reprocessing and adversarial training) and hence could be used in parallel. However, further investigation needs to be done. 

%Our experimental setup is based on simulating a cross-layer implementation from gate-level up to system-level. Therefore, the experiments are  computationally and time demanding, which limited our experiments to the two datasets MNIST and CIFAR-10. This limitation was the same for techniques that require Monte Carlo Simulations such as \cite{snP2019_certif}. Our observations held for 5-layer-CNN (LeNet) and 8-layer-CNN (AlexNet). In future work, we are planning to solve the simulation time limitation by a real hardware implementation, which facilitate evaluating DA for larger networks. 
\textbf{Adaptive attacks:} The main challenge for designing adaptive attacks to DA is to theoretically  model the impact of AC on the gradient, knowing that the approximate multiplication is locally non differentiable (as shown by the noise distribution in Figure \ref{appx_mult}).
However, while AC injects random noise within the model, this noise is not time-variant for the same input. Therefore, we believe that adaptive attacks are possible by defining a surrogate function that models the noisy DNN's gradient. A potential  extension of DA to cope with adaptive attacks is to introducing stochastic approximate noise. The stochastic aspect could be either built-in in the circuit, or injected numerically at higher abstraction levels.

\label{discussion}

\section{Related Work}

%\textbf{Defense strategies.} 
Several defense mechanisms were proposed to combat adversarial attacks and can be categorized as follows:

\noindent
\textbf{Adversarial Training (AT).} 
AT is one of the most explored defenses against adversarial attacks. The main idea can be traced back to \cite{fgsm}, in which models were hardened by including adversarial examples in the training data set of the model. As a result, the trained model classifies evasive samples with higher accuracy. Various attempts to combine AT with other methods have resulted in better defense approaches such as cascade adversarial training \cite{na2017cascade}. %, principled training \cite{certifying}. 
Nonetheless, AT is not effective when the attacker uses a different attack strategy than the one used to train the model~\cite{samangouei2018defense}. Moreover, adversarial training is much more computationally intensive than training a model on the training data set only because generating evasive samples needs more computation and model fitting is more challenging (takes more epochs)~\cite{tramer2017ensemble}.

\noindent
\textbf{Input Preprocessing (IP).}
Input preprocessing depends on applying transformations to the input to remove the adversarial perturbations~\cite{sit, osadchy2017no}. Examples of transformation are %denoising auto-encoders~\cite{gu2014towards}, 
the median, averaging, and Gaussian low-pass filters~\cite{osadchy2017no}, and JPEG compression~\cite{das2017keeping}. However, it was shown that these defenses are insecure under strong white-box attacks~\cite{chen2019towards}; if the attacker knows the specific used transformation, it can be taken into account when creating the attack. Furthermore, preprocessing requires additional computation on every input. 

\noindent
\textbf{Gradient Masking (GM).}
GM relies on applying regularization to the model to make its output less sensitive to input perturbations. Papernot et al. proposed defensive distillation~\cite{distillation_SP}, which is based on increasing the generalization of the model by distilling knowledge out of a large model to train a compact model. Nonetheless, distillation was found weak against $C\&W$ attack \cite{C&W}. %Nayebi and Surya~\cite{nayebi2017biologically} purposed to use saturating networks that use a loss function that promotes the activations to be in their saturating regime. %\cite{ross2018improving} proposed to regularize the gradient input by penalizing variations in the model's output with respect to changes in the input during the training of differentiable models. 
Moreover, GM approaches found to make white-box attacks harder and vulnerable against black-box attacks~\cite{tramer2017ensemble}. Furthermore, they require re-training of pre-trained networks. 

\noindent
\textbf{Randomization-based Defenses.} These techniques are the closest to our work \cite{snP2019_certif,smooth,defense_certified}. Lecuyer et al. \cite{snP2019_certif} also suggest to add random noise to the first layer of the DNN and estimate the output by a Monte Carlo simulation. These techniques offer a bounded theoretical guarantee of robustness. From a practical perspective, none of these works has been evaluated at scale or with realistic implementations.  For example, Raghunathan et al.~\cite{defense_certified} evaluate only a tiny neural network.  Other works~\cite{smooth,snP2019_certif} consider scalability but require high overhead to implement the defense (specifically, to estimate the model output which requires running a heavy Monte Carlo simulation involving a number of different runs of the CNN).
Our approach is different since not only our noise does not require overhead but comes naturally from the simpler and faster AC implementation. Moreover, while these techniques require additional training, DA is a drop-in replacement of the hardware without specific training requirements, and with no changes to the architecture nor the parameters of the CNN.  %In fact, the AC classifier is around 4x faster than the exact classifier, and 2.5x more energy efficient.
\label{sec:rw}

\section{Conclusion}
To the best of our knowledge, this is the first work that proposes the use of hardware-supported approximation as a defense strategy against adversarial attacks for CNNs.  
We propose a CNN implementation based on an energy-efficient approximate floating-point multiplier. While AC is used in the literature to reduce the energy and delay of CNNs, we show that AC also enhances their robustness to adversarial attacks. The proposed defense is, on average, $87\%$ more robust against strong grey-box attacks and $87.5\%$ against strong black-box attacks than a conventional CNN for the case of MNIST dataset, with negligible loss in accuracy.  The approximate CNN achieves a significant reduction in power and delay of $50\%$ to $88\%$ and $67\%$ to $81\%$, respectively. 

%The same approach was used to build an approximate BFloat16 multiplier that resulted in further gain in energy by up to 81\% and 88\% saving in delay.

% if have a single appendix:
%\appendix[Proof of the Zonklar Equations]
% or
%\appendix  % for no appendix heading
% do not use \section anymore after \appendix, only \section*
% is possibly needed

% use appendices with more than one appendix
% then use \section to start each appendix
% you must declare a \section before using any
% \subsection or using \label (\appendices by itself
% starts a section numbered zero.)
%

%\appendices
%\input{theory}

% use section* for acknowledgment
% \ifCLASSOPTIONcompsoc
%   % The Computer Society usually uses the plural form
%   \section*{Acknowledgments}
% \else
%   % regular IEEE prefers the singular form
%   \section*{Acknowledgment}
% \fi

% The authors would like to thank...

% Can use something like this to put references on a page
% by themselves when using endfloat and the captionsoff option.
\ifCLASSOPTIONcaptionsoff
  \newpage
\fi

\begin{IEEEbiography}[{\includegraphics[width=1in,height=1.25in,clip]{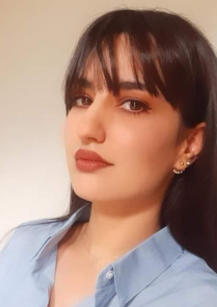}}]{Amira Guesmi} has been pursuing a PhD degree in Computer Systems Engineering at the National Engineering School of Sfax (ENIS), University of Sfax, Tunisia, since 2018. She received the engineer degree in Electrical Engineering from the National Engineering School of Sfax (ENIS), University of Sfax, Tunisia, in 2016. Her research interests include optimizations of deep neural networks in hardware architecture and deep neural network security.
\end{IEEEbiography}
%\vskip -2\baselineskip plus -1fil
\vspace{-29pt}
\begin{IEEEbiography}[{\includegraphics[width=1in,height=1.25in,clip]{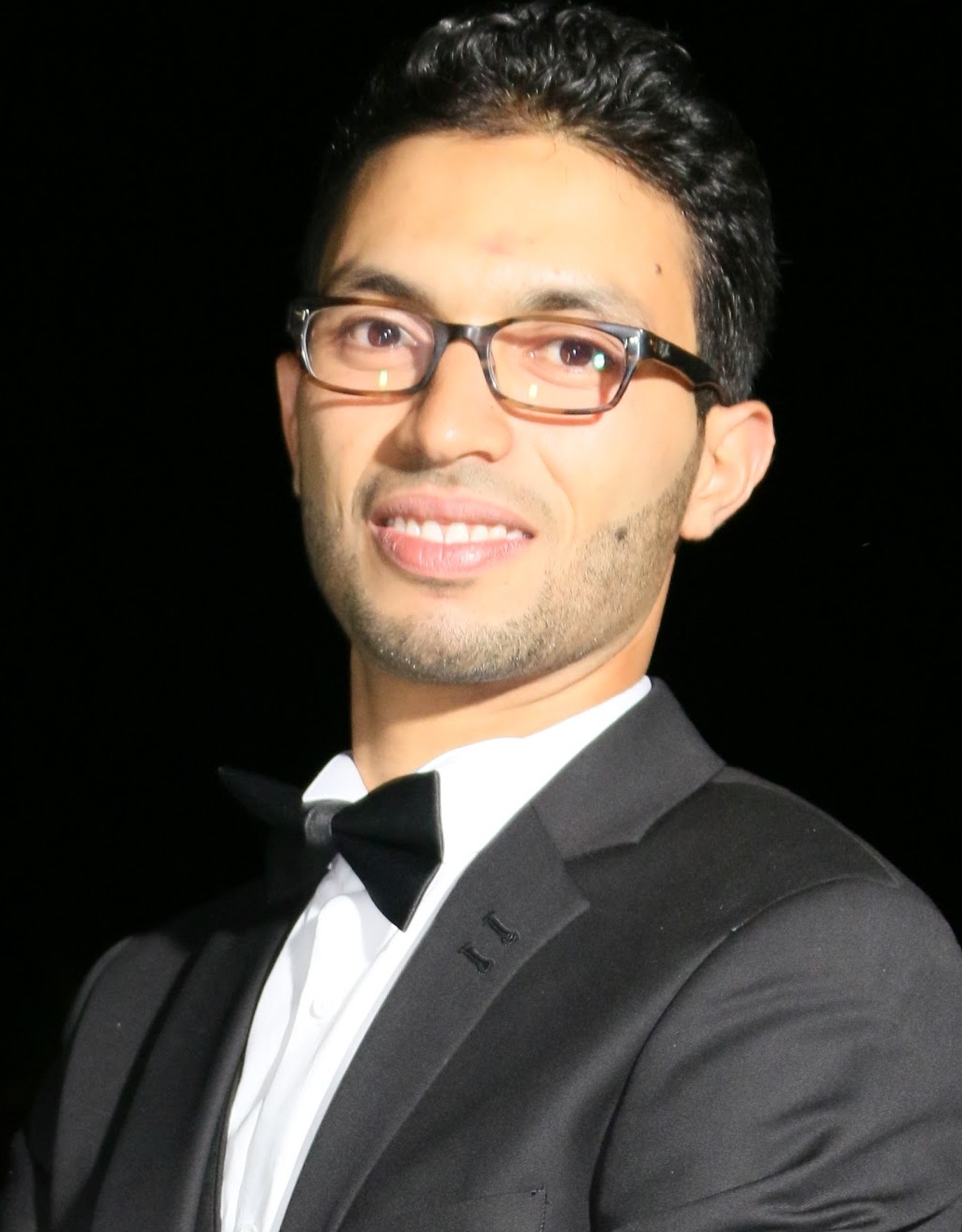}}]{ Ihsen Alouani} is an Associate Professor at the IEMN-DOAE lab (CNRS-8520) and is affiliated to INSA Hauts-De-France. He got his PhD from the Polytechnic University Hauts-De-France, a Msc and engineering degree from the National Engineering School Sousse, Tunisia. His research interest includes Machine Learning, Intelligent Transportation Systems, Data Privacy, Systems Reliability and Security.
\end{IEEEbiography}
\vspace{-29pt}

\begin{IEEEbiography}[{\includegraphics[width=1in,height=1.25in,clip]{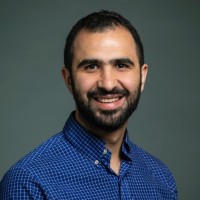}}]{Khaled N. Khasawneh} is an Assistant Professor in the ECE Department at George Mason University. His research interests are in architecture support for security and adversarial machine learning. He received his Ph.D. in Computer Science from the University of California at Riverside in 2019.
\end{IEEEbiography}
\vspace{-29pt}

\begin{IEEEbiography}[{\includegraphics[width=1in,height=1.25in,clip]{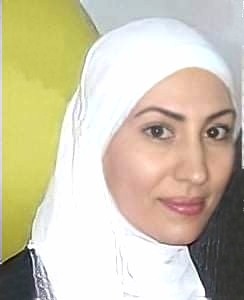}}]{Mouna Baklouti} is currently an Associate Professor in Computer Science at the University of Sfax. She received a HdR in electrical engineering from ENIS School in June 2016. %Since 2012, she has been a co-founder and head of the research master in embedded systems, at the ENIS School in cooperation with the University of Chemnitz, Germany. 
She is a research member of Computer Embedded Systems Laboratory (CES-Lab), (ENIS), Sfax (Tunisia). Her research interests include smart embedded systems design, FPGA-based prototyping, IoT and e-health. She has been an IEEE senior member since 2017.
\end{IEEEbiography}
\vspace{-29pt}

\begin{IEEEbiography}[{\includegraphics[width=1in,height=1.25in,clip]{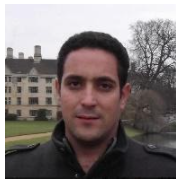}}]{Tarek Frikha} is an assistant professor in National Engineering School of Sfax. %He received the engineer degree in electronic engineering from National School of Engineers of Sfax, in 2006 and the Master Diploma from Polytech Sophia Antipolis in France. 
He received a Ph.D in Science and Technology of Information and Communication in University Of South Brittany, France, and the National School of Engineers of Sfax, Tunisia. His research interests include Multiprocessor architecture optimization for multimedia domains and Hw/Sw codesign, Blockchain for multimedia applications, medical, paramedical, and agricultural applications.
\end{IEEEbiography}
\vspace{-29pt}

\begin{IEEEbiography}[{\includegraphics[width=1in,height=1.25in,clip]{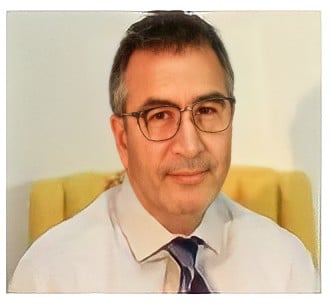}}]{Mohamed Abid} is a Full Professor at the National School of Engineering in Sfax Tunisia. Currently, he is the Director of Computer Embedded System laboratory ‘CES-Lab). He received the Ph. D. degree from the INSAT, Toulouse (France) in 1989 and the ``thèse d’état" degree from the ENIT (Tunisia) in 2000 in the area of Microelectronics. His current research interests include SoC for Embedded Systems. He is an IEEE member. 
\end{IEEEbiography}

\begin{IEEEbiography}[{\includegraphics[width=1in,height=1.25in,clip]{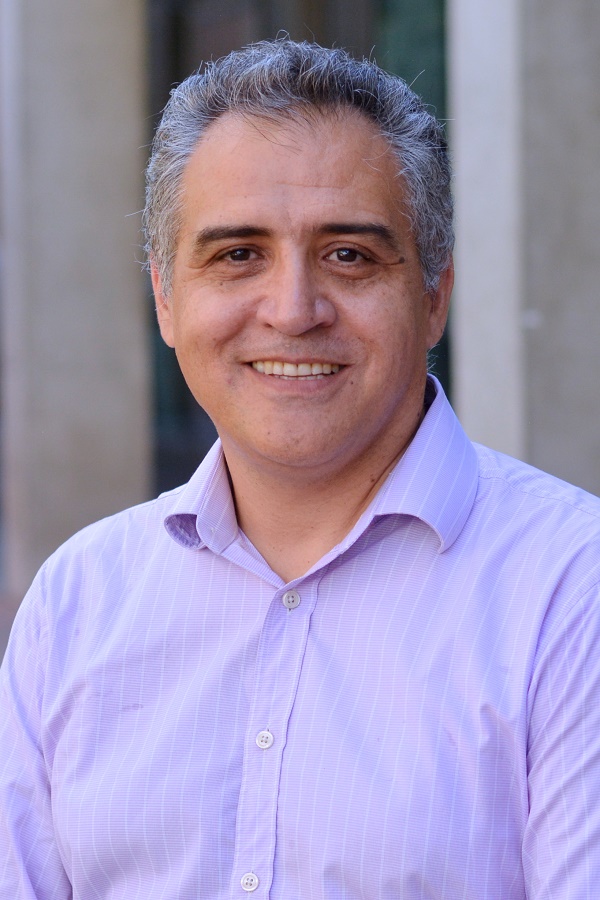}}]{Nael Abu-Ghazaleh} is a Professor with joint appointment in the CSE as well as the ECE departments at the University of California, Riverside, where he also chairs the CE program.  His research interests are broadly in computer systems including architecture security, high performance computing, and networked and distributed systems.  He has published over 200 papers in these areas, several of which have received or been nominated for best paper awards.  His team has identified a number of vulnerabilities that have been reported to major hardware and software companies and influenced commercial products.  He is an ACM distinguished member.
\end{IEEEbiography}

% that's all folks
\end{document}